\documentstyle[12pt]{article}
\newlength{\extraspace}
\newlength{\extraspaces}
\catcode`\@=11
\def\numberbysection{\@addtoreset{equation}{section}
\def\theequation{\arabic{section}.\arabic{equation}}}
\newcommand{\be}{\begin{equation}
\addtolength{\abovedisplayskip}{\extraspaces}
\addtolength{\belowdisplayskip}{\extraspaces}
\addtolength{\abovedisplayshortskip}{\extraspace}
\addtolength{\belowdisplayshortskip}{\extraspace}}
\newcommand{\ee}{\end{equation}}
\newcommand{\ba}{\begin{eqnarray}
\addtolength{\abovedisplayskip}{\extraspaces}
\addtolength{\belowdisplayskip}{\extraspaces}
\addtolength{\abovedisplayshortskip}{\extraspace}
\addtolength{\belowdisplayshortskip}{\extraspace}}
\newcommand{\ea}{\end{eqnarray}}
\newcommand{\newsection}[1]{
\vspace{7mm}
\pagebreak[3]
\addtocounter{section}{1}
\setcounter{equation}{0}
\setcounter{subsection}{0}
\setcounter{footnote}{0}
\begin{center}
{\large {\bf \thesection. #1}}
\end{center}
\nopagebreak
\medskip
\nopagebreak
\hspace{3mm}}
\newcommand{\nonu}{\nonumber \\[.5mm]}
\newcommand{\A}{&\!\!\!}

\newcommand{\e}{\, {\rm e}}
\newcommand{\D}{{\cal D}}

\newcommand{\VEV}[1]{\left\langle {#1} \right\rangle}
\newcommand{\slD}{D \!\!\!\! \raisebox{0.2ex}{/}}
\newcommand{\slp}{p \!\!\!\!\; \raisebox{0.1ex}{/}}
\newcommand{\slpartial}{\partial \!\!\! \raisebox{0.2ex}{/}}

\newcommand{\figone}{
\begin{figure}[tb]
\setlength{\unitlength}{0.0125in}
\begin{picture}(280,60)(-40,770)
\thicklines
\put(100,800){\circle{40}}
\put( 80,800){\line(-1, 0){ 20}}
\put( 60,805){\makebox(0,0)[lb]
    {\raisebox{0pt}[0pt][0pt]{\twlrm $S$}}}
\put(130,795){\makebox(0,0)[lb]
    {\raisebox{1pt}[0pt][0pt]{\twlrm $= 0$}}}
\put(260,800){\circle{40}}
\put(240,800){\line(-1, 0){ 20}}
\put(280,800){\line( 1, 0){ 20}}
\put(220,805){\makebox(0,0)[lb]
    {\raisebox{0pt}[0pt][0pt]{\twlrm $S$}}}
\put(290,805){\makebox(0,0)[lb]
    {\raisebox{0pt}[0pt][0pt]{\twlrm $S$}}}
\put(305,795){\makebox(0,0)[lb]
    {\raisebox{1pt}[0pt][0pt]{\twlrm $= I$}}}
\end{picture}
\caption{The loop diagrams of the scalar field.}
\label{figureone}
\vspace{5mm}
\end{figure}}
\newcommand{\figtwo}{
\begin{figure}[tb]
\setlength{\unitlength}{0.0125in}
\begin{picture}(460,120)(45,630)
\thicklines
\put(185,720){\circle{40}}
\put(165,720){\line(-1, 0){ 20}}
\put(145,725){\makebox(0,0)[lb]
    {\raisebox{0pt}[0pt][0pt]{\twlrm $S$}}}
\put(215,715){\makebox(0,0)[lb]
    {\raisebox{1pt}[0pt][0pt]{\twlrm $= 0$}}}
\put(325,720){\circle{40}}
\put(305,720){\line(-1, 0){ 20}}
\put(285,725){\makebox(0,0)[lb]
    {\raisebox{0pt}[0pt][0pt]{\twlrm $\Omega$}}}
\put(355,715){\makebox(0,0)[lb]
    {\raisebox{1pt}[0pt][0pt]{\twlrm $= 0$}}}
\put(100,660){\circle{40}}
\put( 80,660){\line(-1, 0){ 20}}
\put(120,660){\line( 1, 0){ 20}}
\put( 60,665){\makebox(0,0)[lb]
    {\raisebox{0pt}[0pt][0pt]{\twlrm $S$}}}
\put(130,665){\makebox(0,0)[lb]
    {\raisebox{0pt}[0pt][0pt]{\twlrm $S$}}}
\put(150,655){\makebox(0,0)[lb]{\raisebox{1pt}[0pt][0pt]
    {\twlrm $= \displaystyle{1 \over 2} \, I$}}}
\put(260,660){\circle{40}}
\put(240,660){\line(-1, 0){ 20}}
\put(280,660){\line( 1, 0){ 20}}
\put(220,665){\makebox(0,0)[lb]
    {\raisebox{0pt}[0pt][0pt]{\twlrm $\Omega$}}}
\put(290,665){\makebox(0,0)[lb]
    {\raisebox{0pt}[0pt][0pt]{\twlrm $\Omega$}}}
\put(310,655){\makebox(0,0)[lb]
    {\raisebox{1pt}[0pt][0pt]{\twlrm $= 0$}}}
\put(420,660){\circle{40}}
\put(400,660){\line(-1, 0){ 20}}
\put(440,660){\line( 1, 0){ 20}}
\put(380,665){\makebox(0,0)[lb]
    {\raisebox{0pt}[0pt][0pt]{\twlrm $S$}}}
\put(450,665){\makebox(0,0)[lb]
    {\raisebox{0pt}[0pt][0pt]{\twlrm $\Omega$}}}
\put(470,655){\makebox(0,0)[lb]
    {\raisebox{1pt}[0pt][0pt]{\twlrm $= 0$}}}
\end{picture}
\caption{The loop diagrams of the spinor field.}
\label{figuretwo}
\vspace{5mm}
\end{figure}}
\newcommand{\figthree}{
\begin{figure}[tb]
\setlength{\unitlength}{0.0125in}
\begin{picture}(460,120)(45,490)
\thicklines
\put(180,580){\circle{40}}
\put(160,580){\line(-1, 0){ 20}}
\put(140,585){\makebox(0,0)[lb]
    {\raisebox{0pt}[0pt][0pt]{\twlrm $S$}}}
\put(210,575){\makebox(0,0)[lb]
    {\raisebox{1pt}[0pt][0pt]{\twlrm $= 0$}}}
\put(340,580){\circle{40}}
\put(320,580){\line(-1, 0){ 20}}
\put(300,585){\makebox(0,0)[lb]
    {\raisebox{0pt}[0pt][0pt]{\twlrm $\Omega$}}}
\put(370,575){\makebox(0,0)[lb]
    {\raisebox{1pt}[0pt][0pt]{\twlrm $= 0$}}}
\put(100,520){\circle{40}}
\put( 80,520){\line(-1, 0){ 20}}
\put(120,520){\line( 1, 0){ 20}}
\put( 60,525){\makebox(0,0)[lb]
    {\raisebox{0pt}[0pt][0pt]{\twlrm $S$}}}
\put(130,525){\makebox(0,0)[lb]
    {\raisebox{0pt}[0pt][0pt]{\twlrm $S$}}}
\put(150,515){\makebox(0,0)[lb]
    {\raisebox{1pt}[0pt][0pt]{ $= 12 I$}}}
\put(260,520){\circle{40}}
\put(240,520){\line(-1, 0){ 20}}
\put(280,520){\line( 1, 0){ 20}}
\put(220,525){\makebox(0,0)[lb]
    {\raisebox{0pt}[0pt][0pt]{\twlrm $\Omega$}}}
\put(290,525){\makebox(0,0)[lb]
    {\raisebox{0pt}[0pt][0pt]{\twlrm $\Omega$}}}
\put(310,515){\makebox(0,0)[lb]
    {\raisebox{1pt}[0pt][0pt]{\twlrm $= 6 I$}}}
\put(420,520){\circle{40}}
\put(400,520){\line(-1, 0){ 20}}
\put(440,520){\line( 1, 0){ 20}}
\put(380,525){\makebox(0,0)[lb]
    {\raisebox{0pt}[0pt][0pt]{\twlrm $S$}}}
\put(450,525){\makebox(0,0)[lb]
    {\raisebox{0pt}[0pt][0pt]{\twlrm $\Omega$}}}
\put(470,515){\makebox(0,0)[lb]
   {\raisebox{1pt}[0pt][0pt]{\twlrm $= - 18 I$}}}
\end{picture}
\caption{The $\phi_\alpha$-loop diagrams with vertices from $S_1$.}
\label{figurethree}
\vspace{5mm}
\end{figure}}
\newcommand{\figfour}{
\begin{figure}[tb]
\setlength{\unitlength}{0.0125in}
\begin{picture}(460,180)(45,650)
\thicklines
\put(100,800){\circle{40}}
\put( 80,800){\line(-1, 0){ 20}}
\put( 60,805){\makebox(0,0)[lb]
    {\raisebox{0pt}[0pt][0pt]{\twlrm $S$}}}
\put(130,795){\makebox(0,0)[lb]
    {\raisebox{1pt}[0pt][0pt]{\twlrm $= 0$}}}
\put(260,800){\circle{40}}
\put(240,800){\line(-1, 0){ 20}}
\put(220,805){\makebox(0,0)[lb]
    {\raisebox{0pt}[0pt][0pt]{\twlrm $\Omega$}}}
\put(290,795){\makebox(0,0)[lb]
    {\raisebox{1pt}[0pt][0pt]{\twlrm $= 0$}}}
\put(420,800){\circle{40}}
\put(400,800){\line(-1, 0){ 20}}
\put(380,805){\makebox(0,0)[lb]
    {\raisebox{0pt}[0pt][0pt]{\twlrm $V$}}}
\put(450,795){\makebox(0,0)[lb]
    {\raisebox{1pt}[0pt][0pt]{\twlrm $= 0$}}}
\put(100,740){\circle{40}}
\put( 60,750){\line( 2,-1){ 20}}
\put( 80,740){\line(-2,-1){ 20}}
\put( 60,755){\makebox(0,0)[lb]
    {\raisebox{0pt}[0pt][0pt]{\twlrm $S$}}}
\put( 60,715){\makebox(0,0)[lb]
    {\raisebox{0pt}[0pt][0pt]{\twlrm $S$}}}
\put(130,735){\makebox(0,0)[lb]
    {\raisebox{1pt}[0pt][0pt]{\twlrm $= 3 I$}}}
\put(260,740){\circle{40}}
\put(220,750){\line( 2,-1){ 20}}
\put(240,740){\line(-2,-1){ 20}}
\put(220,755){\makebox(0,0)[lb]
    {\raisebox{0pt}[0pt][0pt]{\twlrm $\Omega$}}}
\put(220,715){\makebox(0,0)[lb]
    {\raisebox{0pt}[0pt][0pt]{\twlrm $\Omega$}}}
\put(290,735){\makebox(0,0)[lb]
    {\raisebox{1pt}[0pt][0pt]{\twlrm $= 0$}}}
\put(420,740){\circle{40}}
\put(380,750){\line( 2,-1){ 20}}
\put(400,740){\line(-2,-1){ 20}}
\put(380,755){\makebox(0,0)[lb]
    {\raisebox{0pt}[0pt][0pt]{\twlrm $V$}}}
\put(380,715){\makebox(0,0)[lb]
    {\raisebox{0pt}[0pt][0pt]{\twlrm $V$}}}
\put(450,735){\makebox(0,0)[lb]
    {\raisebox{1pt}[0pt][0pt]{\twlrm $= - I$}}}
\put(100,680){\circle{40}}
\put( 60,690){\line( 2,-1){ 20}}
\put( 80,680){\line(-2,-1){ 20}}
\put( 60,695){\makebox(0,0)[lb]
    {\raisebox{0pt}[0pt][0pt]{\twlrm $S$}}}
\put( 60,655){\makebox(0,0)[lb]
    {\raisebox{0pt}[0pt][0pt]{\twlrm $\Omega$}}}
\put(130,675){\makebox(0,0)[lb]
    {\raisebox{1pt}[0pt][0pt]{\twlrm $= - 3 I$}}}
\put(260,680){\circle{40}}
\put(220,690){\line( 2,-1){ 20}}
\put(240,680){\line(-2,-1){ 20}}
\put(220,695){\makebox(0,0)[lb]
    {\raisebox{0pt}[0pt][0pt]{\twlrm $\Omega$}}}
\put(220,655){\makebox(0,0)[lb]
    {\raisebox{0pt}[0pt][0pt]{\twlrm $V$}}}
\put(290,675){\makebox(0,0)[lb]
    {\raisebox{1pt}[0pt][0pt]{\twlrm $= 6 I$}}}
\put(420,680){\circle{40}}
\put(380,690){\line( 2,-1){ 20}}
\put(400,680){\line(-2,-1){ 20}}
\put(380,695){\makebox(0,0)[lb]
    {\raisebox{0pt}[0pt][0pt]{\twlrm $V$}}}
\put(380,655){\makebox(0,0)[lb]
    {\raisebox{0pt}[0pt][0pt]{\twlrm $S$}}}
\put(450,675){\makebox(0,0)[lb]
    {\raisebox{1pt}[0pt][0pt]{\twlrm $= - 6 I$}}}
\end{picture}
\caption{The tadpole diagrams with vertices from $S_2$.}
\label{figurefour}
\vspace{5mm}
\end{figure}}
\newcommand{\figfive}{
\begin{figure}[tb]
\setlength{\unitlength}{0.0125in}
\begin{picture}(460,120)(45,710)
\thicklines
\put(100,800){\circle{40}}
\put( 80,800){\line(-1, 0){ 20}}
\put(120,800){\line( 1, 0){ 20}}
\put( 60,805){\makebox(0,0)[lb]
    {\raisebox{0pt}[0pt][0pt]{\twlrm $S$}}}
\put(130,805){\makebox(0,0)[lb]
    {\raisebox{0pt}[0pt][0pt]{\twlrm $S$}}}
\put(150,795){\makebox(0,0)[lb]
    {\raisebox{1pt}[0pt][0pt]{\twlrm $= - 2 I$}}}
\put(260,800){\circle{40}}
\put(240,800){\line(-1, 0){ 20}}
\put(280,800){\line( 1, 0){ 20}}
\put(220,805){\makebox(0,0)[lb]
    {\raisebox{0pt}[0pt][0pt]{\twlrm $\Omega$}}}
\put(290,805){\makebox(0,0)[lb]
    {\raisebox{0pt}[0pt][0pt]{\twlrm $\Omega$}}}
\put(310,795){\makebox(0,0)[lb]
    {\raisebox{1pt}[0pt][0pt]{\twlrm $= 6 I$}}}
\put(420,800){\circle{40}}
\put(400,800){\line(-1, 0){ 20}}
\put(440,800){\line( 1, 0){ 20}}
\put(380,805){\makebox(0,0)[lb]
    {\raisebox{0pt}[0pt][0pt]{\twlrm $V$}}}
\put(450,805){\makebox(0,0)[lb]
    {\raisebox{0pt}[0pt][0pt]{\twlrm $V$}}}
\put(470,795){\makebox(0,0)[lb]
    {\raisebox{1pt}[0pt][0pt]{\twlrm$= \displaystyle{1 \over 2} \, I$}}}
\put(100,740){\circle{40}}
\put( 80,740){\line(-1, 0){ 20}}
\put(120,740){\line( 1, 0){ 20}}
\put( 60,745){\makebox(0,0)[lb]
    {\raisebox{0pt}[0pt][0pt]{\twlrm $S$}}}
\put(130,745){\makebox(0,0)[lb]
    {\raisebox{0pt}[0pt][0pt]{\twlrm $\Omega$}}}
\put(150,735){\makebox(0,0)[lb]
    {\raisebox{1pt}[0pt][0pt]{\twlrm $= - 3 I$}}}
\put(260,740){\circle{40}}
\put(240,740){\line(-1, 0){ 20}}
\put(280,740){\line( 1, 0){ 20}}
\put(220,745){\makebox(0,0)[lb]
    {\raisebox{0pt}[0pt][0pt]{\twlrm $\Omega$ }}}
\put(290,745){\makebox(0,0)[lb]
    {\raisebox{0pt}[0pt][0pt]{\twlrm $V$}}}
\put(310,735){\makebox(0,0)[lb]
    {\raisebox{1pt}[0pt][0pt]{\twlrm $= - 6 I$}}}
\put(420,740){\circle{40}}
\put(400,740){\line(-1, 0){ 20}}
\put(440,740){\line( 1, 0){ 20}}
\put(380,745){\makebox(0,0)[lb]
    {\raisebox{0pt}[0pt][0pt]{\twlrm $V$}}}
\put(450,745){\makebox(0,0)[lb]
    {\raisebox{0pt}[0pt][0pt]{\twlrm $S$}}}
\put(470,735){\makebox(0,0)[lb]
    {\raisebox{1pt}[0pt][0pt]{\twlrm $= 6 I$}}}
\end{picture}
\caption{The $\phi_\alpha$-loop diagrams with two vertices from $S_2$.}
\label{figurefive}
\vspace{5mm}
\end{figure}}
\newcommand{\figsix}{
\begin{figure}[tb]
\setlength{\unitlength}{0.0125in}
\begin{picture}(460,120)(45,570)
\thicklines
\put(100,660){\circle{40}}
\put( 80,660){\line(-1, 0){ 20}}
\put(120,660){\line( 1, 0){ 20}}
\put( 60,665){\makebox(0,0)[lb]
    {\raisebox{0pt}[0pt][0pt]{\twlrm $S$}}}
\put(130,665){\makebox(0,0)[lb]
    {\raisebox{0pt}[0pt][0pt]{\twlrm $S$}}}
\put(150,655){\makebox(0,0)[lb]
    {\raisebox{1pt}[0pt][0pt]{\twlrm $= 0$}}}
\put(260,660){\circle{40}}
\put(240,660){\line(-1, 0){ 20}}
\put(280,660){\line( 1, 0){ 20}}
\put(220,665){\makebox(0,0)[lb]
    {\raisebox{0pt}[0pt][0pt]{\twlrm $\Omega$}}}
\put(290,665){\makebox(0,0)[lb]
    {\raisebox{0pt}[0pt][0pt]{\twlrm $\Omega$}}}
\put(310,655){\makebox(0,0)[lb]
    {\raisebox{1pt}[0pt][0pt]{\twlrm $= 0$}}}
\put(420,660){\circle{40}}
\put(400,660){\line(-1, 0){ 20}}
\put(440,660){\line( 1, 0){ 20}}
\put(380,665){\makebox(0,0)[lb]
    {\raisebox{0pt}[0pt][0pt]{\twlrm $S$}}}
\put(450,665){\makebox(0,0)[lb]
    {\raisebox{0pt}[0pt][0pt]{\twlrm $V$}}}
\put(470,655){\makebox(0,0)[lb]
    {\raisebox{1pt}[0pt][0pt]{\twlrm $= 0$}}}
\put(100,600){\circle{40}}
\put( 80,600){\line(-1, 0){ 20}}
\put(120,600){\line( 1, 0){ 20}}
\put( 60,605){\makebox(0,0)[lb]
    {\raisebox{0pt}[0pt][0pt]{\twlrm $S$}}}
\put(130,605){\makebox(0,0)[lb]
    {\raisebox{0pt}[0pt][0pt]{\twlrm $\Omega$}}}
\put(150,595){\makebox(0,0)[lb]
    {\raisebox{1pt}[0pt][0pt]{\twlrm $= - 6 I$}}}
\put(260,600){\circle{40}}
\put(240,600){\line(-1, 0){ 20}}
\put(280,600){\line( 1, 0){ 20}}
\put(220,605){\makebox(0,0)[lb]
    {\raisebox{0pt}[0pt][0pt]{\twlrm $\Omega$}}}
\put(290,605){\makebox(0,0)[lb]
    {\raisebox{0pt}[0pt][0pt]{\twlrm $S$}}}
\put(310,595){\makebox(0,0)[lb]
    {\raisebox{1pt}[0pt][0pt]{\twlrm $= - 6 I$}}}
\put(420,600){\circle{40}}
\put(400,600){\line(-1, 0){ 20}}
\put(440,600){\line( 1, 0){ 20}}
\put(380,605){\makebox(0,0)[lb]
    {\raisebox{0pt}[0pt][0pt]{\twlrm $\Omega$}}}
\put(450,605){\makebox(0,0)[lb]
    {\raisebox{0pt}[0pt][0pt]{\twlrm $V$}}}
\put(470,595){\makebox(0,0)[lb]
    {\raisebox{1pt}[0pt][0pt]{\twlrm $= 12 I$}}}
\end{picture}
\caption{The $\phi_\alpha$-loop diagrams with one vertex from $S_1$
(left vertex) and the other from $S_2$ (right vertex).}
\label{figuresix}
\vspace{5mm}
\end{figure}}
\newcommand{\figseven}{
\begin{figure}[tb]
\setlength{\unitlength}{0.0125in}
\begin{picture}(460,180)(45,650)
\thicklines
\put(100,800){\circle{40}}
\put( 80,800){\line(-1, 0){ 20}}
\put( 60,805){\makebox(0,0)[lb]
    {\raisebox{0pt}[0pt][0pt]{\twlrm $S$}}}
\put(130,795){\makebox(0,0)[lb]
    {\raisebox{1pt}[0pt][0pt]{\twlrm $= 0$}}}
\put(260,800){\circle{40}}
\put(240,800){\line(-1, 0){ 20}}
\put(220,805){\makebox(0,0)[lb]
    {\raisebox{0pt}[0pt][0pt]{\twlrm $\Omega$}}}
\put(290,795){\makebox(0,0)[lb]
    {\raisebox{1pt}[0pt][0pt]{\twlrm $= 0$}}}
\put(420,800){\circle{40}}
\put(400,800){\line(-1, 0){ 20}}
\put(380,805){\makebox(0,0)[lb]
    {\raisebox{0pt}[0pt][0pt]{\twlrm $\hat R$}}}
\put(450,795){\makebox(0,0)[lb]
    {\raisebox{1pt}[0pt][0pt]{\twlrm $= 6 I$}}}
\put(200,740){\circle{40}}
\put(160,750){\line( 2,-1){ 20}}
\put(180,740){\line(-2,-1){ 20}}
\put(160,755){\makebox(0,0)[lb]
    {\raisebox{0pt}[0pt][0pt]{\twlrm $\Omega$}}}
\put(160,715){\makebox(0,0)[lb]
    {\raisebox{0pt}[0pt][0pt]{\twlrm $\Omega$}}}
\put(235,735){\makebox(0,0)[lb]
    {\raisebox{1pt}[0pt][0pt]{\twlrm +}}}
\put(300,740){\circle{40}}
\put(280,740){\makebox(0.4444,0.6667){\tenrm .}}
\put(280,740){\line(-1, 0){ 20}}
\put(320,740){\line( 1, 0){ 20}}
\put(260,745){\makebox(0,0)[lb]
    {\raisebox{0pt}[0pt][0pt]{\twlrm $\Omega$}}}
\put(330,745){\makebox(0,0)[lb]
    {\raisebox{0pt}[0pt][0pt]{\twlrm $\Omega$}}}
\put(350,735){\makebox(0,0)[lb]
    {\raisebox{1pt}[0pt][0pt]{\twlrm $= 0$}}}
\put(100,680){\circle{40}}
\put( 80,680){\makebox(0.4444,0.6667){\tenrm .}}
\put( 80,680){\line(-1, 0){ 20}}
\put(120,680){\line( 1, 0){ 20}}
\put( 60,685){\makebox(0,0)[lb]
    {\raisebox{0pt}[0pt][0pt]{\twlrm $S$}}}
\put(130,685){\makebox(0,0)[lb]
    {\raisebox{0pt}[0pt][0pt]{\twlrm $S$}}}
\put(150,675){\makebox(0,0)[lb]
    {\raisebox{1pt}[0pt][0pt]{\twlrm $= 4 I$}}}
\put(260,680){\circle{40}}
\put(240,680){\makebox(0.4444,0.6667){\tenrm .}}
\put(240,680){\line(-1, 0){ 20}}
\put(280,680){\line( 1, 0){ 20}}
\put(220,685){\makebox(0,0)[lb]
    {\raisebox{0pt}[0pt][0pt]{\twlrm $S$ }}}
\put(290,685){\makebox(0,0)[lb]
    {\raisebox{0pt}[0pt][0pt]{\twlrm $\Omega$}}}
\put(310,675){\makebox(0,0)[lb]
    {\raisebox{1pt}[0pt][0pt]{\twlrm $= 0$}}}
\put(420,680){\circle{40}}
\put(400,680){\line(-1, 0){ 20}}
\put(440,680){\line( 1, 0){ 20}}
\put(380,685){\makebox(0,0)[lb]
    {\raisebox{0pt}[0pt][0pt]{\twlrm $S$}}}
\put(450,685){\makebox(0,0)[lb]
    {\raisebox{0pt}[0pt][0pt]{\twlrm $\hat R$}}}
\put(470,675){\makebox(0,0)[lb]
    {\raisebox{1pt}[0pt][0pt]{\twlrm $= 0$}}}
\end{picture}
\caption{The loop diagrams of the supersymmetry ghosts
$\beta$ and $\gamma$.}
\label{figureseven}
\vspace{5mm}
\end{figure}}
\newcommand{\figeight}{
\begin{figure}[tb]
\setlength{\unitlength}{0.0125in}
\begin{picture}(460,120)(45,710)
\thicklines
\put(100,800){\circle{40}}
\put( 80,800){\line(-1, 0){ 20}}
\put( 60,805){\makebox(0,0)[lb]
    {\raisebox{0pt}[0pt][0pt]{\twlrm $S$}}}
\put(130,795){\makebox(0,0)[lb]
    {\raisebox{1pt}[0pt][0pt]{\twlrm $= 0$}}}
\put(215,800){\circle{40}}
\put(195,800){\line(-1, 0){ 20}}
\put(175,805){\makebox(0,0)[lb]
    {\raisebox{0pt}[0pt][0pt]{\twlrm $\Omega$}}}
\put(245,795){\makebox(0,0)[lb]
    {\raisebox{1pt}[0pt][0pt]{\twlrm $= 0$}}}
\put(330,800){\circle{40}}
\put(310,800){\line(-1, 0){ 20}}
\put(290,805){\makebox(0,0)[lb]
    {\raisebox{0pt}[0pt][0pt]{\twlrm $T$}}}
\put(357.5,795){\makebox(0,0)[lb]
    {\raisebox{1pt}[0pt][0pt]{\twlrm +}}}
\put(415,800){\circle{40}}
\put(395,800){\line(-1, 0){ 20}}
\put(435,800){\line( 1, 0){ 20}}
\put(375,805){\makebox(0,0)[lb]
    {\raisebox{0pt}[0pt][0pt]{\twlrm $\Omega$}}}
\put(445,805){\makebox(0,0)[lb]
    {\raisebox{0pt}[0pt][0pt]{\twlrm $\Omega$}}}
\put(465,795){\makebox(0,0)[lb]
    {\raisebox{1pt}[0pt][0pt]{\twlrm $= - 12 I$}}}
\put(100,740){\circle{40}}
\put( 80,740){\line(-1, 0){ 20}}
\put( 60,745){\makebox(0,0)[lb]
    {\raisebox{0pt}[0pt][0pt]{\twlrm $\hat R$}}}
\put(130,735){\makebox(0,0)[lb]
    {\raisebox{1pt}[0pt][0pt]{\twlrm $= 0$}}}
\put(260,740){\circle{40}}
\put(240,740){\line(-1, 0){ 20}}
\put(280,740){\line( 1, 0){ 20}}
\put(220,745){\makebox(0,0)[lb]
    {\raisebox{0pt}[0pt][0pt]{\twlrm $S$}}}
\put(290,745){\makebox(0,0)[lb]
    {\raisebox{0pt}[0pt][0pt]{\twlrm $\Omega$}}}
\put(310,735){\makebox(0,0)[lb]
    {\raisebox{1pt}[0pt][0pt]{\twlrm $= 0$}}}
\put(420,740){\circle{40}}
\put(400,740){\line(-1, 0){ 20}}
\put(440,740){\line( 1, 0){ 20}}
\put(380,745){\makebox(0,0)[lb]
    {\raisebox{0pt}[0pt][0pt]{\twlrm $S$}}}
\put(450,745){\makebox(0,0)[lb]
    {\raisebox{0pt}[0pt][0pt]{\twlrm $S$}}}
\put(470,735){\makebox(0,0)[lb]
    {\raisebox{1pt}[0pt][0pt]{\twlrm $= 3 I$}}}
\end{picture}
\caption{The graviton ($h_{\alpha\beta}$, $\phi$) loop diagrams.}
\label{figureeight}
\vspace{5mm}
\end{figure}}
\newcommand{\fignine}{
\begin{figure}[tb]
\setlength{\unitlength}{0.0125in}
\begin{picture}(460,120)(45,510)
\thicklines
\put(100,600){\circle{40}}
\put( 80,600){\line(-1, 0){ 20}}
\put( 60,605){\makebox(0,0)[lb]
    {\raisebox{0pt}[0pt][0pt]{\twlrm $S$}}}
\put(130,595){\makebox(0,0)[lb]
    {\raisebox{1pt}[0pt][0pt]{\twlrm $= 0$}}}
\put(260,600){\circle{40}}
\put(240,600){\line(-1, 0){ 20}}
\put(220,605){\makebox(0,0)[lb]
{\raisebox{0pt}[0pt][0pt]{\twlrm $\Omega$}}}
\put(290,595){\makebox(0,0)[lb]
    {\raisebox{1pt}[0pt][0pt]{\twlrm $= 0$}}}
\put(420,600){\circle{40}}
\put(400,600){\line(-1, 0){ 20}}
\put(440,600){\line( 1, 0){ 20}}
\put(380,605){\makebox(0,0)[lb]
    {\raisebox{0pt}[0pt][0pt]{\twlrm $S$}}}
\put(450,605){\makebox(0,0)[lb]
    {\raisebox{0pt}[0pt][0pt]{\twlrm $\Omega$}}}
\put(470,595){\makebox(0,0)[lb]
    {\raisebox{1pt}[0pt][0pt]{\twlrm $= 0$}}}
\put(120,540){\circle{40}}
\put(100,540){\line(-1, 0){ 20}}
\put( 80,545){\makebox(0,0)[lb]
    {\raisebox{0pt}[0pt][0pt]{\twlrm $T$}}}
\put(155,535){\makebox(0,0)[lb]
    {\raisebox{1pt}[0pt][0pt]{\twlrm +}}}
\put(220,540){\circle{40}}
\put(200,540){\line(-1, 0){ 20}}
\put(240,540){\line( 1, 0){ 20}}
\put(180,545){\makebox(0,0)[lb]
{\raisebox{0pt}[0pt][0pt]{\twlrm $\Omega$}}}
\put(250,545){\makebox(0,0)[lb]
{\raisebox{0pt}[0pt][0pt]{\twlrm $\Omega$}}}
\put(270,535){\makebox(0,0)[lb]
    {\raisebox{1pt}[0pt][0pt]{\twlrm $= - 12 I$}}}
\put(400,540){\circle{40}}
\put(380,540){\line(-1, 0){ 20}}
\put(420,540){\line( 1, 0){ 20}}
\put(360,545){\makebox(0,0)[lb]
    {\raisebox{0pt}[0pt][0pt]{\twlrm $S$}}}
\put(430,545){\makebox(0,0)[lb]
    {\raisebox{0pt}[0pt][0pt]{\twlrm $S$}}}
\put(450,535){\makebox(0,0)[lb]
    {\raisebox{1pt}[0pt][0pt]{\twlrm $= - 4 I$}}}
\end{picture}
\caption{The loop diagrams of the ghosts $b$ and $c$ for
general coordinate symmetry.}
\label{figurenine}
\vspace{5mm}
\end{figure}}
\newcommand{\figten}{
\begin{figure}[tb]
\setlength{\unitlength}{0.0125in}
\begin{picture}(460,60)(65,510)
\thicklines
\put(140,540){\circle{40}}
\put(120,540){\line(-1, 0){ 20}}
\put(100,545){\makebox(0,0)[lb]
    {\raisebox{0pt}[0pt][0pt]{\twlrm $T'$}}}
\put(175,535){\makebox(0,0)[lb]
    {\raisebox{1pt}[0pt][0pt]{\twlrm +}}}
\put(240,540){\circle{40}}
\put(220,540){\line(-1, 0){ 20}}
\put(260,540){\line( 1, 0){ 20}}
\put(200,545){\makebox(0,0)[lb]
    {\raisebox{0pt}[0pt][0pt]{\twlrm $\Omega$}}}
\put(270,545){\makebox(0,0)[lb]
    {\raisebox{0pt}[0pt][0pt]{\twlrm $\Omega$}}}
\put(290,535){\makebox(0,0)[lb]
    {\raisebox{1pt}[0pt][0pt]{\twlrm $= - 6 I$}}}
\put(420,540){\circle{40}}
\put(400,540){\line(-1, 0){ 20}}
\put(440,540){\line( 1, 0){ 20}}
\put(380,545){\makebox(0,0)[lb]
    {\raisebox{0pt}[0pt][0pt]{\twlrm $S$}}}
\put(450,545){\makebox(0,0)[lb]
    {\raisebox{0pt}[0pt][0pt]{\twlrm $S$}}}
\put(470,535){\makebox(0,0)[lb]
    {\raisebox{1pt}[0pt][0pt]{\twlrm $= 2 I$}}}
\end{picture}
\caption{The $A_\alpha$-loop diagrams.}
\label{figureten}
\vspace{5mm}
\end{figure}}
\begin{document}
\addtolength{\baselineskip}{.7mm}
\thispagestyle{empty}
\begin{flushright}
TIT/HEP--238 \\
Imperial/TP/93--94/8 \\
{\tt hep-th/9311045} \\
November, 1993
\end{flushright}
\vspace{2mm}
\begin{center}
{\Large{\bf Supergravity in $2+\epsilon$ Dimensions}} \\[15mm]
{\sc Shin-ichi Kojima},\footnote{
\tt e-mail: kotori@phys.titech.ac.jp} \hspace{5mm}
{\sc Norisuke Sakai}\footnote{
\tt e-mail: nsakai@phys.titech.ac.jp} \\[3mm]
{\it Department of Physics, Tokyo Institute of Technology \\[2mm]
Oh-okayama, Meguro, Tokyo 152, Japan} \\[4mm]
and \\[4mm]
{\sc Yoshiaki Tanii}\footnote{
On leave of absence from Physics Department,
Saitama University, Urawa, Saitama 338, Japan. \\
\hspace*{6.5mm}{\tt e-mail: y.tanii@ic.ac.uk}} \\[3mm]
{\it The Blackett Laboratory, Imperial College \\[2mm]
London, SW7 2BZ, U.K.} \\[15mm]
{\bf Abstract}\\[5mm]
{\parbox{13cm}{\hspace{5mm}
Supergravity theory in $2+\epsilon$ dimensions is studied.
It is invariant under supertransformations in $2$ and $3$ dimensions.
One-loop divergence is explicitly computed in the background field
method and a nontrivial fixed point is found.
In quantizing the supergravity, a gauge fixing condition is devised
which explicitly isolates conformal and superconformal modes.
The renormalization of the gravitationally dressed operators is
studied and their anomalous dimensions are computed.
Problems to use the dimensional reduction are also examined.
}}
\end{center}
\vfill
\newpage
\setcounter{section}{0}
\setcounter{equation}{0}
\numberbysection
%
%%%%%%%  Section 1  %%%%%%%%%%%%%%%%%%%%%%%%%%%%%%%%%%%%%%%%
%
\newsection{Introduction}
Quantum theory of gravity has been an outstanding challenge for many
years. It is power-counting renormalizable at two spacetime dimensions.
It has been proposed to study the quantum theory
of gravity at $d=2+\epsilon$ dimensions and to expand it in powers
of $\epsilon$ \cite{WEI}--\cite{GK}.
After paying due attention to subtleties associated with
a conformal mode,
a nontrivial fixed point has been found at one-loop order \cite{KN}.
More recently, renormalization properties of the $2+\epsilon$
dimensional quantum gravity have been further studied using a
convenient choice of variables and gauge fixing conditions \cite{KKN}.
\par
Quantum gravity in two spacetime dimensions is useful not only as a
theoretical laboratory for higher dimensional gravity theories,
but also as a basis for string theory.
Two-dimensional supergravity is especially important for string
theories, although it
is also interesting from the point of view of quantum gravity theories.
Recent progress in matrix models provides possibilities for a
nonperturbative treatment of the two-dimensional quantum
gravity \cite{GRMI}.
(For a review of matrix models, see ref.\ \cite{GIMO}.)
However, so far it has been difficult to incorporate supersymmetry
in the two-dimensional spacetime in such a discretized approach.
At the moment we need to develop continuum approaches
to study supergravity in two dimensions.
Therefore it is useful to explore a computational scheme to deal with
the supergravity theory at and near two-dimensional spacetime,
in order to understand the superstring theory as well.
\par
The purpose of this paper is to formulate a supergravity theory in
$d=2+\epsilon$ dimensions and to compute the beta function up to
one-loop order.
We find a nontrivial fixed point of order $\epsilon$, analogous to the
nonsupersymmetric case.
We also study the $\epsilon \rightarrow 0$ limit of the theory and
find that the result is in agreement with the usual continuum approach
using conformal field theory \cite{DDK}--\cite{GX}.
\par
We shall present an action for supergravity multiplet and supersymmetric
matter multiplet in $d=2+\epsilon$ dimensions, which is exactly
invariant under the supersymmetry transformation at $2$ and $3$
dimensions.
Our theory smoothly interpolates between $d=2$ and $d=3$ supergravity
theories \cite{DZBVH}--\cite{HUY}, but is not exactly invariant at
noninteger dimensions.
This is primarily because the proof of invariance requires Fierz
identities which are valid only at integer dimensions.
We shall attempt to use dimensional reduction \cite{SIEGELDR}
from $D$ to $d$ dimensions, for instance, $D=3$ to $d=2+\epsilon$.
However, we find that the proof of invariance is plagued by
inconsistencies which are similar to those encountered in the case
of $D=4$ dimensions \cite{SIEGELIC}.
\par
We shall use the background field method \cite{DEWITT}, \cite{TV}
to compute one-loop counter terms.
We introduce a two-parameter family of gauge fixing conditions
which is convenient to
separate conformal and superconformal modes from the rest of the
supergravity multiplet.
By choosing the gauge parameters, we can obtain a gauge in which
(super)conformal modes have no mixing with the
non(super)conformal modes, and moreover the propagators
of non(super)conformal modes have no $1/\epsilon$ pole.
This gauge choice facilitates computations of one-loop counter terms
significantly.
This gauge can be understood as a supersymmetric generalization of
the gauge adopted for the $2+\epsilon$ dimensional gravity \cite{KKN}.
\par
One-loop counter terms are obtained by computing
contributions from the superconformal
modes (spin $1/2$ components) and those from the non-superconformal
modes (spin $3/2$ components) of the gravitino field separately.
They are combined with contributions from superghosts and
Nakanishi-Lautrup fields.
The contributions from the graviton sector
are also computed and found to agree with refs.\ \cite{KN}, \cite{KKN}.
By combining all these one-loop counter terms,
we obtain beta function for the
gravitational coupling constant $G$.
The resulting beta function turns out to have a nontrivial fixed point
at
\be
G^* = {\epsilon \over 9-\hat c},
\ee
where $\hat c$ denotes the central charge of the superconformal
matter multiplet such as $\hat c$ free scalar and spinor fields.
By taking the $\epsilon \rightarrow 0$ limit, we can define a
two-dimensional quantum gravity theory.
We shall construct physical operators in such a limit and compute their
scaling dimensions.
The results are found to agree with the conformal field theory
approach \cite{DDK}--\cite{GX}.
\par
In the next section, supergravity theory in $d=2+\epsilon$ dimensions
is presented together with supersymmetric matter field theories.
In sect.\ 3, the supergravity theory is quantized by introducing
gauge fixing conditions which separate
(super)conformal and non(super)conformal modes.
One-loop divergences are computed in sect.\ 4.
In sect.\ 5, a nontrivial fixed point is shown to follow from the
beta function and the two-dimensional limit is also worked out.
In sect.\ 6, dimensional reduction from $D$ to $d$ dimensions
is examined and is shown to possess inconsistencies if we want to
use it to prove invariance under supersymmetry transformations.
In appendix A one-loop counterterms of a vector gauge field are
computed. In appendices B and C details of dimensional reductions
of supergravity action are worked out.
%
%%%%%%%  Section 2  %%%%%%%%%%%%%%%%%%%%%%%%%%%%%%%%%%%%%%%%
%
\newsection{Supergravity in $2+\epsilon$ dimensions}
The $N = 1$ supergravity multiplet in two- and
three-dimensional spacetime
consists of a vielbein $e_\mu{}^\alpha$,
a Majorana Rarita-Schwinger field $\psi_{\mu}$ and a real
scalar auxiliary field $S$ \cite{DZBVH}--\cite{HUY}.
The local Lorentz indices are denoted by Greek letters starting
$\alpha, \beta, \cdots $,
and the world indices are denoted by middle Greek letters
starting $\mu, \nu, \cdots $.
Both indices run from $0$ to $d-1$.
A signature of the metric is $(-, +, \cdots, +)$.
We shall use the same field content as a supergravity multiplet
in $d = 2+\epsilon$ dimensions.
The action in $d=2+\epsilon$ dimensions is an interpolation
between two- and three-dimensional ones \cite{HBG}, \cite{HUY}
\be
S_{\rm SG}={1 \over 16\pi G_0}\int d^d x \, e \left[ R
+ i \bar\psi_{\mu} \gamma^{\mu\nu\rho} D_{\nu} \psi_{\rho}
- {d-2 \over d-1} S^2 \right],
\label{supergravityaction}
\ee
where $e = \det e_{\mu}{}^\alpha$ and $G_0$ is the bare
gravitational coupling constant.
The multi-index matrices $\gamma^{\mu\nu\cdots}$ are the
antisymmetrized products of gamma
matrices such as $\gamma^{\mu\nu\rho}
={1 \over 3!}(\gamma^{\mu}\gamma^{\nu}\gamma^{\rho}
\pm \mbox{permutations})$.
The scalar curvature in the Einstein term is given by
\be
R = e_\alpha{}^\mu e_\beta{}^\nu
\left( \partial_\mu \omega_\nu{}^{\alpha\beta}
    + \omega_\mu{}^\alpha{}_\gamma \, \omega_\nu{}^{\gamma\beta}
    - (\mu \leftrightarrow \nu) \right).
\ee
Let us note that the kinetic term of the gravitino $\psi_\rho$ is
invariant under the general coordinate transformation
without the affine connection (Christoffel symbol $\Gamma$) in
the covariant derivative $D_{\nu}$
\be
 D_{\nu} \psi_{\rho} = \left( \partial_{\nu} +
{1 \over 4}\omega_{\nu}{}^{\alpha\beta}
\gamma_{\alpha\beta} \right) \psi_{\rho}.
\ee
In the spirit of the so-called second order formalism
\cite{VANNIEUVEN}, the spin connection
$\omega_\mu{}^{\alpha\beta}$ is chosen such that it satisfies the
equation of motion derived from the above action
\be
D_\mu e_\nu{}^\alpha - D_\nu e_\mu{}^\alpha
= - {1 \over 2} i \bar\psi_\mu \gamma^\alpha \psi_\nu
\equiv T_{\mu\nu}{}^\alpha,
\label{torsioneq}
\ee
where $T_{\mu\nu}{}^\alpha$ is a torsion.
Again there is no affine connection in the covariant derivatives.
The solution of the above equation of motion for the spin connection is
given by
\be
\omega_\mu{}^{\alpha\beta}
= \omega_\mu{}^{\alpha\beta}(e) + \kappa_\mu{}^{\alpha\beta},
\label{spinconnection}
\ee
where $\omega_\mu{}^{\alpha\beta}(e)$ is the usual spin connection
without torsion
\be
\omega_{\mu\alpha\beta}
= {1 \over 2} e_\alpha{}^\rho e_\beta{}^\sigma \Bigl[ \,
e_\sigma{}^\gamma \left( \partial_\mu e_{\rho\gamma}
- \partial_\rho e_{\mu\gamma} \right)
- e_\mu{}^\gamma \partial_\rho e_{\sigma\gamma}
- (\rho \leftrightarrow \sigma) \, \Bigr]
\ee
and the torsion part is explicitly separated as
\be
\kappa_{\mu \alpha\beta} = - {1 \over 4} \, i
\left( \bar\psi_\mu \gamma_\alpha \psi_\beta
- \bar\psi_\mu \gamma_\beta \psi_\alpha
+ \bar\psi_\alpha \gamma_\mu \psi_\beta \right).
\label{torsionpart}
\ee
\par
The supertransformations of the supergravity multiplet are given
in terms of an infinitesimal anticommuting  parameter $\varepsilon(x)$
\ba
\delta e_\mu{}^\alpha
\A = \A - i \bar\varepsilon \gamma^\alpha \psi_{\mu}, \nonu
\delta \psi_{\mu} \A = \A 2 \left(D_{\mu} +
{1 \over 2(d-1)} \, S \, \gamma_{\mu}\right)\varepsilon, \nonu
\delta S \A = \A {1 \over 2} i S \, \bar\varepsilon
\gamma^{\mu} \psi_{\mu}
- {1 \over 2} i \bar\varepsilon \gamma^{\mu\nu} \psi_{\mu\nu},
\label{supertrans}
\ea
where $\psi_{\mu\nu}$ is the antisymmetrized covariant derivative
of the Rarita-Schwinger field without the affine connection
\be
\psi_{\mu\nu} = D_{\mu}\psi_{\nu}- D_{\nu}\psi_{\mu}.
\ee
The transformation of the spin connection turns out to be
\be
\delta \omega_{\mu \alpha\beta}
= {1 \over 2} i ( \bar\varepsilon \gamma_\alpha \psi_{\mu \beta}
  - \bar\varepsilon \gamma_\beta \psi_{\mu \alpha}
  + \bar\varepsilon \gamma_\mu \psi_{\alpha\beta} )
  + {1 \over 2(d-1)} i S ( \bar\varepsilon \psi_\beta e_{\mu \alpha}
  - \bar\varepsilon \psi_\alpha e_{\mu \beta}
  + \bar\varepsilon \gamma_{\alpha\beta} \psi_\mu ).
\label{super_spin_conn}
\ee
We do not need this transformation of the spin connection
to prove the invariance of the supergravity action
since we have chosen the spin connection such that it
satisfies the equation of motion.
\par
Let us show the invariance of the supergravity action
(\ref{supergravityaction})
for $d=2$ and $d=3$ \cite{HBG}, \cite{HUY}.
For $d=2$ it is trivial since
the action is vanishing. To prove the invariance for $d=3$, it is
convenient to rewrite the Rarita-Schwinger action as
\be
S_{\rm RS} = - {1 \over 16\pi G_0}\int d^d x \, i \,
\epsilon^{\mu\nu\rho} \, \bar\psi_\mu D_\nu \psi_\rho.
\ee
Under the supertransformation, each term in
the action (\ref{supergravityaction}) is transformed into
\ba
\delta S_{\rm E}
\A = \A {1 \over 16 \pi G_0}
\int d^3 x \, e \, (-i) \, \bar\varepsilon \gamma^\alpha \psi_\mu
\left( R \, e_\alpha{}^\mu - 2 R_\alpha{}^\mu \right), \nonu
\delta S_{\rm RS}
\A = \A {1 \over 16 \pi G_0}
\int d^3 x \, e \left[ \, i \bar\varepsilon \gamma^\alpha \psi_\mu
\left( R \, e_\alpha{}^\mu - 2 R_\alpha{}^\mu \right)
- {1 \over 2} \, i S \bar\varepsilon \gamma^{\mu\nu} \psi_{\mu\nu}
\, \right], \nonu
\delta S_{\rm aux}
\A = \A {1 \over 16 \pi G_0} \int d^3 x \, e \, {1 \over 2} \,
i S \, \bar\varepsilon \gamma^{\mu\nu} \psi_{\mu\nu}.
\ea
We see that the supergravity action is invariant under the
supertransformation.
For $d \not= 2, 3$, we can show that the action
(\ref{supergravityaction}) is invariant under the
supertransformations (\ref{supertrans}) neglecting terms of order
$(\mbox{fermi fields})^3$.
\par
As a matter multiplet in $N=1$ supergravity theory we take
a scalar supermultiplet consisting of a real scalar field
$X$, a Majorana spinor field $\lambda$ and a real scalar
auxiliary filed $F$. The matter action is an interpolation between
the two- and three-dimensional ones \cite{DZBVH}--\cite{HUY}
\ba
S_{\rm M} \A = \A \int d^d x \, e \biggl[ -{1 \over 2} g^{\mu\nu}
       \partial_\mu X \partial_\nu X
       + {1 \over 2} i \bar\lambda \gamma^\mu D_\mu \lambda
       + {1 \over 2} F F \nonu
\A \A  + {1 \over 2} i \bar\psi_\mu \gamma^\nu \gamma^\mu
       \lambda \partial_\nu X
       - {d-2 \over 4(d-1)} i S \bar\lambda \lambda
       - {1 \over 16} \bar\psi_\mu \gamma^\nu \gamma^\mu
       \psi_\nu \bar\lambda \lambda \biggr],
\label{matteraction}
\ea
where the spin connection contains the torsion part
$\kappa_{\mu\alpha\beta}$
as is defined in eq.\ (\ref{torsionpart}).
The matter supermultiplet is noninteracting if we switch off
the gravitational interactions in the flat spacetime limit.
We shall consider $\hat c$ of such matter supermultiplets:
$(X^i, \lambda^i, F^i)$  $(i = 1, \cdots, \hat c)$.
\par
Supertransformations for the matter multiplet are given by
\ba
\delta X \A = \A i \bar\varepsilon \lambda, \nonu
\delta \lambda \A = \A - \gamma^\mu \varepsilon D^P_\mu X
                    - \varepsilon F, \nonu
\delta F \A = \A i \bar\varepsilon \gamma^\mu D^P_\mu \lambda
           - {d-2 \over 2(d-1)} i S \bar\varepsilon \lambda,
\label{supertrmatter}
\ea
where $D^P_\mu$ is the supercovariant derivative:
\ba
D^P_\mu X \A = \A \partial_\mu X
- {1 \over 2} i \bar\psi_\mu \lambda, \nonu
D^P_\mu \lambda \A = \A D_\mu \lambda
+ {1 \over 2} \gamma^\nu \psi_\mu D^P_\nu X
+ {1 \over 2} \psi_\mu F.
\label{supercovariantder}
\ea
Supertransformation of the matter action gives residual terms
which are of order $(\mbox{fermi fields})^3$ and vanish
for $d = 2, 3$ once we use the Fierz identities.
The Fierz identities in two and three dimensions are
\ba
d = 2 \A : \A \quad
\bar\chi_1 \chi_2 \bar\chi_3 \chi_4
= - {1 \over 2} \Bigl[
\bar\chi_1 \chi_4 \bar\chi_3 \chi_2
+ \bar\chi_1 \gamma^\alpha \chi_4
\bar\chi_3 \gamma_\alpha \chi_2
- {1 \over 2} \bar\chi_1 \gamma^{\alpha\beta} \chi_4
\bar\chi_3 \gamma_{\alpha\beta} \chi_2 \Bigr], \nonu
d = 3 \A : \A \quad
\bar\chi_1 \chi_2 \bar\chi_3 \chi_4
= - {1 \over 2} \Bigl[
\bar\chi_1 \chi_4 \bar\chi_3 \chi_2
+ \bar\chi_1 \gamma^\alpha \chi_4
\bar\chi_3 \gamma_\alpha \chi_2 \Bigr],
\label{fierz}
\ea
where $\chi_1, \cdots, \chi_4$ are arbitrary anticommuting
spinors.
Let us note that the Fierz identities imply that we are assuming
spinors to be two-component and the gamma
matrices $\gamma^\alpha$ to be $2 \times 2$ matrices.
In proving the invariance of the matter action
it is necessary to use the transformation of the spin connection
(\ref{super_spin_conn}). The order $(\mbox{fermi field})^3$ terms
remain for $d \not = 2, 3$.
\par
By using a superfield formalism \cite{HBG}, \cite{DHK},
we can construct physical operators in two dimensions,
which are invariant under all the local gauge symmetries.
We consider only the simplest operator $O_p$, which corresponds to
the tachyon vertex operator with momentum $p$
in the Neveu-Schwarz sector of string theories.
In terms of the component fields it is given by
\ba
O_p \A = \A \int d^d x \, O_p(x), \nonu
O_p(x) \A = \A e \left( i p_i \bar\lambda^i \lambda^j p_j
      - 2 i p_i F^i + \bar\psi_\mu \gamma^\mu
      \lambda^i p_i - 2 S + {1 \over 2} i \bar\psi_\mu
      \gamma^{\mu\nu} \psi_\nu \right) e^{i p \cdot X}.
\label{physicalop}
\ea
We now consider this operator in general $d$ dimensions.
By using the gamma matrix identities in two or three dimensions,
we find the supertransformation of the integrand
\be
\delta O_p(x)
= \partial_\mu \left[ 2 e ( \bar\varepsilon \gamma^\mu
  \lambda^i p_i + i \bar\varepsilon \gamma^{\mu\nu} \psi_\nu )
  e^{i p \cdot X} \right].
\ee
Therefore $O_p$ is invariant not only in two dimensions but also
in three dimensions.
For three dimensions, we have used the following identity
\be
\epsilon^{\mu\nu\rho} \bar\varepsilon \gamma^\alpha
\psi_\rho \bar\psi_\mu \gamma_\alpha \psi_\nu = 0,
\ee
which is valid because of the Fierz identity.
For $d \not= 2, 3$, we can show that the operator (\ref{physicalop})
is invariant under the supertransformations neglecting terms of order
$(\mbox{fermi fields})^3$.
The above operator can be considered as a bare operator and
we need to consider its renormalization properties. In particular
we should consider the dressing due to the conformal mode.
We shall describe the case of the two-dimensional limit in detail in
sect.\ 5.
%
%%%%%%%  Section 3  %%%%%%%%%%%%%%%%%%%%%%%%%%%%%%%%%%%%%%%%
%
\newsection{Gauge fixing and quantization}
To find one-loop counterterms we shall compute one-loop divergences
using the action (\ref{supergravityaction}) with the bare
gravitational constant $G_0$ replaced by $G / \mu^\epsilon$,
where $G$ and $\mu$ are the renormalized gravitational constant
and the renormalization scale respectively.
We use the background field method \cite{DEWITT}, \cite{TV}.
Fields $\Phi$ are written as a
sum of background fields $\hat\Phi$ and quantum fields
$\Phi_q$: $\Phi = \hat\Phi + \Phi_q$.
At one-loop level, the effective action $\Gamma[\hat\Phi]$, i.e.,
the generating functional of one-particle-irreducible diagrams
is given by a path integral
\be
\e^{i \Gamma[\hat\Phi]}
= \e^{i S[\hat\Phi]}
  \int \D \Phi_q \, \e^{i S^{(2)} [ \Phi_q; \, \hat\Phi ]},
\label{effaction}
\ee
where $S^{(2)}$ is a part of the action which is quadratic
in the quantum fields $\Phi_q$.
\par
We shall denote the background vielbein as $\hat e_\mu{}^\alpha$
and use a
parametrization for quantum fields to separate conformal mode
(trace part) $\phi$ from nonconformal mode (traceless part)
$h_\alpha{}^\beta$ \cite{KKN}
\be
e_\mu{}^\alpha
=  \hat e_\mu{}^\beta (\e^{{1 \over 2}\kappa h} )_\beta{}^\alpha
               \e^{-{1 \over 2}\kappa\phi}, \quad
h_{\alpha\beta} \equiv h_\alpha{}^\gamma \eta_{\gamma\beta}
= h_{\beta\alpha}, \quad
h_\alpha{}^\alpha = 0,
\label{conformalcond}
\ee
where $\kappa^2 = 16\pi G / \mu^\epsilon$.
We have fixed the local Lorentz symmetry such that $h_{\alpha\beta}$
is symmetric. The field $\phi$ is called Liouville field.
For the Rarita-Schwinger field,
we also introduce a parametrization to separate superconformal mode
(spin $1/2$ part) $\eta$ from the nonsuperconformal mode
(spin $3/2$ part) $\phi_\alpha$
\be
\psi_\mu = \kappa \, e_\mu{}^\alpha (\phi_\alpha + \gamma_\alpha \eta),
\qquad
 \gamma^\alpha \phi_\alpha = 0.
\label{superconformal}
\ee
We have chosen background fields other than that of the
vielbein to vanish.
In terms of these parametrizations, a part of the supergravity
action which is quadratic in the quantum fields is given by
\ba
S^{(2)}_{\rm SG} \A = \A
\int d^d x \, \hat{e} \Biggl[ \,
- {1 \over 4} \hat D_\mu h_{\alpha\beta} \hat D^\mu h^{\alpha\beta}
+ {1 \over 4}(d-2)(d-1)\hat{g}^{\mu\nu} \partial_\mu \phi
\partial_\nu \phi \nonu
\A \A + {1 \over 2} \hat D_\alpha h^{\alpha\gamma}
\hat D_\beta h^\beta{}_\gamma
- {1 \over 2} h^{\alpha\beta} h^{\gamma\delta}
\hat R_{\alpha\gamma\delta\beta}
+ {1 \over 8}(d-2)^2 \phi^2 \hat R \nonu
\A \A + {1 \over 2} (d-2) \phi h^{\alpha\beta} \hat R_{\alpha\beta}
- {1 \over 2}(d-2)\phi \hat{D}_\alpha \hat{D}_\beta h^{\alpha\beta}
+ \, i \, \bar\phi^\alpha \hat\slD \, \phi_\alpha \nonu
\A \A + (d-2) \, i \left( \bar\eta \hat{D}_\alpha \phi^\alpha
- \bar\phi^\alpha \hat{D}_\alpha \eta \right)
- (d-2)(d-1) \, i \, \bar\eta \hat\slD \, \eta
- {d-2 \over d-1} S^2 \, \Biggr],
\label{quadaction}
\ea
where $\hat\slD=\gamma^\alpha \hat D_\alpha$ and we have rescaled
the auxiliary field $S$ by $\kappa$.
\par
To fix the gauge symmetries of general coordinate
and local supersymmetry transformations, we use the method
of ref.\ \cite{KU}. In this method the gauge symmetry is
fixed by adding a BRST exact term $- i \delta_{\rm B} (b F)$ to the
lagrangian,
where $b$ is the anti-ghost field and $F$ is a gauge function.
The gauge function $F$ is an arbitrary function of the fields.
This method is equivalent to the procedure of refs.\ \cite{KANI},
\cite{VANNIEUVEN}.
At one-loop level we can discuss the gauge fixing of general
coordinate symmetry and local supersymmetry separately.
\par
To fix general coordinate symmetry we introduce
fermionic Faddeev-Popov ghost and anti-ghost fields $c^\alpha, b_\alpha$
and a bosonic Nakanishi-Lautrup auxiliary field $B_\alpha$,
all of which are
vector fields.
The BRST transformations of $h_{\alpha\beta}$, $\phi$,
$b_\alpha$ and $B_\alpha$ are given by
\ba
\delta_{\rm B} h_{\alpha\beta} \A = \A
\hat D_\alpha c_\beta + \hat D_\beta c_\alpha
-{2 \over d}\eta_{\alpha\beta} \hat D_\gamma c^\gamma + \cdots, \nonu
\delta_{\rm B} \phi \A = \A - {2 \over d} \hat D_\alpha c^\alpha
+ \cdots, \nonu
\delta_{\rm B} b_\alpha \A = \A i B_\alpha, \qquad
\delta_{\rm B} B_\alpha = 0,
\label{gcbrst}
\ea
where the dots represent terms quadratic and higher order in
the quantum fields.
We will use the following gauge function
with two parameters $\alpha$, $\beta$ to fix the general coordinate
symmetry\footnote{Our gauge parameters can be compared
to those used in other papers
such as ref.\ \cite{KN} which is denoted by a suffix KN:
$\beta^{\rm ours}=\beta^{\rm KN} d - 2$,
$\alpha^{\rm ours}=1/\alpha^{\rm KN}$.}
\be
(F_{\rm GC})_\alpha
= \hat e \left( \hat D^\beta h_{\beta\alpha}
         + {1 \over 2} \beta  \hat D_\alpha \phi
         + {1 \over 2} \alpha B_\alpha \right).
\label{gcfunc}
\ee
Then the gauge fixing term and the ghost action are given by
\ba
S_{\rm GC}
\A = \A
\int d^d x \, \delta_{\rm B} \bigl[ - i
b^\alpha (F_{\rm GC})_\alpha \bigr] \nonu
\A = \A
 \int d^d x \, \hat e \, \Biggl[ \,
{1 \over 2} \alpha B'^\alpha B'_\alpha
- {1 \over 2\alpha}
  \left( \hat D^\beta h_{\alpha\beta}
+ {\beta  \over 2}\hat D_\alpha \phi \right)^2 \nonu
\A \A + i  b^\alpha
\left( \hat D^\beta ( \hat D_\alpha c_\beta + \hat D_\beta c_\alpha)
- {\beta +2 \over d}\hat D_\alpha \hat D_\beta c^\beta \right)
+ \cdots \, \Biggr],
\label{gcgf}
\ea
where $B'_\alpha$ is a shifted auxiliary field.
\par
For local supersymmetry gauge fixing, we introduce
bosonic Faddeev-Popov ghost and antighost fields $\gamma$, $\beta$,
and a fermionic Nakanishi-Lautrup field $B$, all of which are
spinor fields. The BRST transformations are
\ba
\delta_{\rm B} \phi_\alpha \A = \A
{2 \over d} \Bigl( (d-1) \eta_{\alpha\beta} -
\gamma_{\alpha\beta} \Bigr) \hat D^\beta \gamma + \cdots, \nonu
\delta_{\rm B} \eta
\A = \A {2 \over d} \hat\slD \; \gamma + \cdots, \nonu
\delta_{\rm B} \beta \A = \A B, \qquad
\delta_{\rm B} B = 0.
\label{susybrst}
\ea
We use the gauge fixing condition with two parameters
$a$ and $b$
\be
F_{\rm SUSY}
= 2 \hat e \left( \hat D^\alpha \phi_\alpha
  - \hat\slD \left( b \eta - {a \over 2} B \right) \right).
\label{susyfunc}
\ee
The gauge fixing term and the ghost action are
\ba
S_{\rm SUSY} \A = \A
\int d^d x \, \delta_{\rm B} \bigl[ - i \bar\beta F_{\rm SUSY} \bigr]
\nonu
\A = \A \int d^d x \, i \, \hat e \,
\biggl[ \,
- {1 \over a}\bar{\phi}^\alpha \hat{D}_\alpha \hat\slD^{-1}
\hat{D}^\beta \phi_\beta
- {b \over a} (\bar\eta \hat D_\alpha \phi^\alpha
- \bar{\phi}^\alpha \hat D_\alpha \eta)
+ {b^2 \over a}\bar\eta \hat\slD \; \eta \nonu
\A \A - {4 \over d} \, \bar\beta \left( (d-1) \hat D^\alpha \hat D_\alpha
- \gamma^{\alpha\beta} \hat D_\alpha \hat D_\beta
- b \hat\slD^2 \right) \gamma
- a \bar B' \hat\slD \, B' + \cdots \, \biggr],
\label{susygf}
\ea
where we have shifted the Nakanishi-Lautrup field to eliminate mixing
with other fields
\be
B' = B +{1 \over a}(\hat\slD^{-1}\hat D_\alpha \phi^\alpha - b \eta).
\ee
In our gauge condition, the Nakanishi-Lautrup field
$B'$ is free but is propagating unlike ordinary auxiliary fields,
whereas the nonsuperconformal mode (spin $3/2$ part) $\phi_\alpha$
acquires a nonlocal term.
\par
So far we treated the gauge fixing for the general coordinate
symmetry and the local supersymmetry separately.
Since the gauge fixing condition for one symmetry can actually
violate the other symmetry too, we need to consider the combined
BRST transformation of the total gauge conditions
$- i \delta_{\rm B} ( b^\alpha (F_{\rm GC})_\alpha
+ \bar\beta F_{\rm SUSY} )$
as the gauge fixing and ghost terms, where the BRST transformation
refers to a combined BRST transformations of (\ref{gcbrst}) and
(\ref{susybrst}).
The resulting cross terms do not contribute
   at one-loop level, since they are higher orders in quantum fields.
After the gauge fixing, the quadratic part of the total
action is
\ba
S^{(2)}_{\rm tot}
\A = \A S^{(2)}_{\rm SG} + S^{(2)}_{\rm GC} + S^{(2)}_{\rm SUSY} \nonu
\A = \A \int d^d x \, \hat e \, \biggl[ \,
- {1 \over 4} \hat D_\mu h_{\alpha\beta} \hat D^\mu h^{\alpha\beta}
+ {1 \over 4} \left( (d-2)(d-1) - {\beta^2 \over 2\alpha} \right)
\hat{g}^{\mu\nu} \partial_\mu \phi \partial_\nu \phi \nonu
\A \A
+ {\alpha - 1 \over 2\alpha}
\hat D_\alpha h^{\alpha\gamma} \hat D_\beta h^{\beta}{}_\gamma
- {1 \over 2} h^{\alpha\beta} h^{\gamma\delta}
\hat R_{\alpha\gamma\delta\beta}
+ {1 \over 8} (d-2)^2 \phi^2 \hat R \nonu
\A \A + {1 \over 2} (d-2) \phi h^{\alpha\beta} \hat R_{\alpha\beta}
- {1 \over 2} \left( d - 2 - {\beta \over \alpha} \right)
\phi \hat{D}_\alpha \hat{D}_\beta h^{\alpha\beta} \nonu
\A \A
+ \, i \, \bar\phi^\alpha \hat\slD \; \phi_\alpha
- {1 \over a} \, i \, \bar{\phi}^\alpha \hat{D}_\alpha
\hat\slD^{-1} \hat{D}^\beta \phi_\beta
- \left( (d-2)(d-1) - {b^2 \over a} \right) \, i \,
  \bar\eta \hat\slD \; \eta \nonu
\A \A
+ \left( d - 2 - {b \over a} \right) \, i \,
  (\bar\eta \hat{D}_\alpha \phi^\alpha
- \bar\phi^\alpha \hat{D}_\alpha \eta)
+ {1 \over 2} \alpha B'^\alpha B'_\alpha
- a \, i \, \bar B' \hat\slD \; B' \nonu
\A \A -{d-2 \over d-1}S^2
+ \mbox{ the ghost terms } \biggr].
\label{totaction}
\ea
\par
We choose the gauge parameters such that propagators have
no mixing
\be
\VEV{h_{\alpha\beta}(x) \phi(y)} = 0, \qquad
\VEV{\phi_\alpha(x) \bar\eta(y)} = 0.
\ee
{}From eq.\ (\ref{totaction})
we find that these conditions require
\be
\beta =  (d-2) \alpha , \qquad
b =  (d-2) a.
\label{nomix}
\ee
It is most convenient to use propagators in the flat background
metric by separating the background metric into the flat metric and
the fluctuations $\hat h_\alpha{}^\beta$, $\hat\phi$ as
\be
\hat e_\mu{}^\alpha= \delta_\mu{}^\beta
( \e^{{1 \over 2} \hat h} )_\beta{}^\alpha \e^{-{1 \over 2} \hat\phi}
= \delta_\mu{}^\alpha + {1 \over 2} \hat h_\mu{}^\alpha
- {1 \over 2} \delta_\mu{}^\alpha \hat\phi + \cdots.
\label{backgroundexp}
\ee
When the relations (\ref{nomix}) are satisfied, the propagators
on the flat background metric are given by
\ba
\VEV{h_{\alpha\beta}(x) h_{\gamma\delta}(y)}
\A = \A - i \int {d^d p \over (2\pi)^d}
\biggl[\eta_{\alpha\gamma}\eta_{\beta\delta}
+\eta_{\alpha\delta}\eta_{\beta\gamma}
-{2(2-\alpha) \over 2+2\epsilon-\alpha\epsilon}
\eta_{\alpha\beta}\eta_{\gamma\delta} \nonu
\A \A - (1-\alpha)
{\eta_{\alpha\gamma} p_\beta p_\delta
+ \eta_{\beta\delta} p_\alpha p_\gamma
+ \eta_{\alpha\delta} p_\beta p_\gamma
+ \eta_{\beta\gamma} p_\alpha p_\delta \over p^2} \nonu
\A \A + {4(1-\alpha) \over 2+2\epsilon-\alpha\epsilon}
{\eta_{\alpha\beta} p_\gamma p_\delta
+ \eta_{\gamma\delta} p_\alpha p_\beta \over  p^2} \nonu
\A \A + {4\epsilon(1-\alpha)^2 \over 2+2\epsilon-\alpha\epsilon}
{p_\alpha p_\beta p_\gamma p_\delta \over  p^4}
\biggr] {1 \over p^2} \e^{i p \cdot (x-y)}, \nonu
\VEV{\phi(x) \phi(y)}
\A = \A i \int {d^d p \over (2\pi)^d}
{4 \over \epsilon(2+2\epsilon-\alpha\epsilon)} {1 \over p^2}
\e^{i p \cdot (x-y)},
\label{grprop} \\[.5mm]
\VEV{\phi_\alpha(x) \bar\phi_\beta(y)}
\A = \A - {1 \over 2(1 + \epsilon - a \epsilon)} \, i
  \int {d^d p \over (2\pi)^d}
  \biggl[  (1-a) \gamma_{\alpha\gamma\beta} p^\gamma
+ a (\gamma_\alpha p_\beta + \gamma_\beta p_\alpha ) \nonu
\A \A + (a +(1-a)\epsilon) \eta_{\alpha\beta} \slp
  - (4a+\epsilon) {p_\alpha \slp p_\beta \over p^2}
\biggr]
  {1 \over p^2} \e^{i p \cdot (x-y)}, \nonu
\VEV{\eta(x) \bar\eta(y)}
\A = \A  {1 \over 2 \epsilon ( 1 + \epsilon - a \epsilon )} \, i
  \int {d^d p \over (2\pi)^d} \, {\slp \over p^2}
  \e^{i p \cdot (x-y)}.
\label{rsprop}
\ea
We see that the propagators of $\phi$ and $\eta$ have a factor
$\epsilon^{-1}$.
This is due to the vanishing of the supergravity action in $d = 2$.
In addition, there are gauge parameter dependent factors such as
$1/(2+2\epsilon -\alpha\epsilon)$.
They can become singular, since the gauge parameters can depend on
$\epsilon$.
It is convenient to choose gauge parameters
such that these factors do not diverge in the limit
$\epsilon \rightarrow 0$.
This is achieved by choosing $\alpha$ and $a$ finite
in the limit $\epsilon \rightarrow 0$.
The propagators (\ref{grprop}) in the gravity sector
are simplified by choosing $\alpha = 1, \beta = \epsilon $.
This is the gauge used in ref.\ \cite{KKN}.
The Rarita-Schwinger propagators (\ref{rsprop}) are simplified by
choosing $a = 1,\ b = \epsilon$. We will use this gauge
to compute one-loop divergences of the Rarita-Schwinger
field in the next section.
\par
The gauge function of the local supersymmetry (\ref{susyfunc})
can be guessed from the supertransformation of the gauge function
(\ref{gcfunc}) of the general coordinate symmetry.
It is useful to consider a rigid supersymmetry in flat space
by taking the transformation parameter $\varepsilon$ in
eq.\ (\ref{supertrans}) to be covariantly constant
$D_{\mu}\varepsilon=0$ and to supplement
with the transformation for the Nakanishi-Lautrup fields
$\delta B_\alpha = - i \bar\varepsilon \partial_\alpha B$.
Then we find that these two gauge functions are related each other,
under the following identification of the gauge parameters
\be
a = {\alpha \over d+2}, \qquad
b = {\beta  \over d+2}.
\label{susyrel}
\ee
The conditions (\ref{nomix}) are consistent with these identifications.
%
%%%%%%%  Section 4  %%%%%%%%%%%%%%%%%%%%%%%%%%%%%%%%%%%%%%%%
%
\newsection{One-loop divergences}
Let us first illustrate our methods taking the matter supermultiplet
as an example.
The action for the matter multiplet in eq.\ (\ref{matteraction})
can be expanded in powers of the quantum fields.
In order to compute the one-loop divergences from the matter loop,
we only need to consider quadratic terms in propagating quantum fields.
By the general covariance for the background field and the
dimensional analysis, divergences in the effective action
(\ref{effaction}) are proportional to the
Einstein action. To compute the coefficients we
expand the background field in powers of fluctuation on the
flat metric as in eq.\ (\ref{backgroundexp}).
We define a unit of divergence as
\ba
I \A = \A - {1 \over 24\pi \epsilon} \int d^d x \, \hat e \hat R \nonu
\A = \A - {1 \over 24\pi \epsilon} \int d^d x
\left[ -{1 \over 4}\partial_{\mu}\hat h_{\nu\rho}
\partial^{\mu}\hat h^{\nu\rho}+\cdots \right],
\label{unitanomaly}
\ea
where we have explicitly exhibited a term in the right-hand side
which survives when we impose conditions
\be
\partial_\mu \hat h^{\mu\nu} = 0, \qquad
\hat h^{\mu\nu} = \hat h^{\nu\mu}, \qquad
\hat \phi = 0.
\label{transvtraceless}
\ee
The coefficients of divergences can be obtained by computing
terms quadratic in $\hat h_{\mu\nu}$ which satisfies
eq.\ (\ref{transvtraceless}).
These conditions simplify the loop calculations in many cases.
\par
For the scalar field in the matter supermultiplet,
we consider the following decomposition
of the action quadratic in the quantum field $X$
\ba
S^{(2)}_{\rm scalar} \A = \A
-{1 \over 2} \int d^d x \, \hat e \hat g^{\mu\nu}
       \partial_\mu X \partial_\nu X \nonu
\A = \A - {1 \over 2} \int d^d x  \left[ \eta^{\mu\nu}
       \partial_\mu X \partial_\nu X
 + S^{\mu\nu} \partial_\mu X \partial_\nu X \right].
\ea
The first term gives the propagator on the flat background, and the
second term gives an interaction with background metric through
the vertex $S^{\mu\nu}$
\figone
\be
S^{\mu\nu} = \hat e \hat g^{\mu\nu} -\eta^{\mu\nu}
= -\hat h^{\mu\nu} + O(\hat h^2),
\label{svertex}
\ee
where we have used the conditions (\ref{transvtraceless}).
As shown in Fig.\ \ref{figureone},
the tadpole diagram does not diverge at one-loop order,
whereas the diagram with two $S$ vertices
gives the following one-loop divergence
in the effective action $\Gamma$ defined in (\ref{effaction})
\be
\Gamma_{\rm scalar} = 1 \times I.
\label{scalaranomaly}
\ee
\par
Similarly the action quadratic in the Majorana spinor field can be
decomposed into free part and interaction vertices
\ba
S^{(2)}_{\rm spinor} \A = \A
{1 \over 2}\int d^d x \, \hat e \, i \bar\lambda \hat \slD \lambda
\nonu
\A=\A
{1 \over 2} \int d^d x \, i \left[ \bar\lambda \slpartial \lambda
+ S_{\alpha}{}^{\mu} \bar \lambda \gamma^{\alpha} \partial_{\mu} \lambda
+ {1 \over 4} \Omega_{\alpha\beta\gamma} \bar \lambda \gamma^{\alpha}
\gamma^{\beta\gamma} \lambda \right],
\ea
\be
 S_{\alpha}{}^{\mu} =
\hat e \hat e_{\alpha}{}^{\mu}-\delta_{\alpha}^{\mu}
= - {1 \over 2} \, \hat h_{\alpha}{}^{\mu} + O(\hat h^2), \qquad
 \Omega_{\alpha\beta\gamma} = \hat e \hat e_{\alpha}{}^{\mu}
\hat \omega_{\mu\beta\gamma}.
\ee
\figtwo
As shown in Fig.\ \ref{figuretwo}, only the diagram with
two $S$ vertices gives the
one-loop divergence
\be
\Gamma_{\rm spinor} = {1 \over 2} \times I.
\label{spinoranomaly}
\ee
As another example of our methods, we compute one-loop
divergences of a vector gauge field in appendix A.
\par
Next let us consider contributions given by the Rarita-Schwinger
field. For simplicity, we choose $a=1, b=\epsilon$ gauge.
The quadratic part of the action for the Rarita-Schwinger field with
the gauge fixing term and the ghost term
can be decomposed into free part and interaction vertices
\ba
S^{(2)}_{\rm RS} + S^{(2)}_{\rm SUSY}
\A = \A \int d^d x \, i
\left[ \bar \phi^{\alpha} \slpartial \phi_{\alpha}
+ \partial_{\alpha} \bar\phi^{\alpha}
{1 \over \slpartial}\partial_{\beta} \phi^{\beta}
- \epsilon \, \hat e \, \bar\eta \hat\slD \eta \right]
+ S_1 + S_2 + S_{{\rm SUSY-FP}}, \label{susyrs} \nonu
\A \A \\[.5mm]
S_1 \A = \A \int d^d x \, i \left[
S_{\beta}{}^\mu \bar\phi^{\alpha} \gamma^{\beta}
\partial_\mu \phi_{\alpha}
+ {1 \over 4}\Omega_{\beta\gamma\delta} \bar\phi^{\alpha}
\gamma^\beta \gamma^{\gamma\delta} \phi_{\alpha}
+ \Omega_{\beta\alpha\gamma} \bar\phi^{\alpha} \gamma^{\beta}
\phi^{\gamma} \right], \label{phivertex1} \nonu
\A \A \\[.5mm]
S_2 \A = \A \int d^d x \, i \Biggl[
\left( S_{\alpha}{}^\mu \partial_\mu \bar\phi^\alpha
- {1 \over 4}\Omega_{\alpha\gamma\delta}
\bar\phi^{\alpha} \gamma^{\gamma\delta}
+ \Omega_\alpha{}^{\alpha\gamma} \bar\phi_\gamma \right) \!
\left( {1 \over \slpartial} - {1 \over \slpartial}
V {1 \over \slpartial}
\right) \! \partial_{\beta} \phi^{\beta} \nonu
\A \A + \partial_{\beta} \bar\phi^{\beta} \!
\left( {1 \over \slpartial}
  - {1 \over \slpartial}V{1 \over \slpartial} \right) \!
  \left( S_{\alpha}{}^\mu \partial_\mu \phi^\alpha
+ {1 \over 4} \Omega_{\alpha\gamma\delta}
  \gamma^{\gamma\delta} \phi^{\alpha}
+ \Omega_\alpha{}^{\alpha\gamma} \phi_\gamma \right) \nonu
\A \A + \left( S_\alpha{}^\mu \partial_\mu \bar\phi^\alpha
  - {1 \over 4}\Omega_{\alpha\gamma\delta} \bar\phi^\alpha
\gamma^{\gamma\delta} \right)
  {1 \over \slpartial} \left( S_\beta{}^\nu \partial_\nu \phi^\beta
   + {1 \over 4}\Omega_{\beta\epsilon\zeta} \gamma^{\epsilon\zeta}
\phi^\beta \right) \nonu
\A \A - \partial_\alpha \bar\phi^\alpha \left( {1 \over \slpartial}
V{1 \over \slpartial}
  - {1 \over \slpartial}V{1 \over \slpartial}V{1 \over \slpartial}
\right)
  \partial_\beta \phi^\beta \Biggr],
\label{phivertex2} \\[.5mm]
S_{{\rm SUSY-FP}}
\A = \A \int d^d x \, i \biggl[ \,
{4 \over d} \Bigl(
\eta^{\mu\nu} \partial_\mu \bar\beta \partial_\nu \gamma
+ S^{\mu\nu} \partial_\mu \bar\beta \partial_\nu \gamma \nonu
\A \A + {1 \over 4}\Omega^{\mu \alpha\beta} (\partial_\mu \bar\beta
\gamma_{\alpha\beta} \gamma
  - \bar\beta \gamma_{\alpha\beta} \partial_\mu \gamma )
-  {1 \over 16}\Omega_{\mu}{}^{\alpha\beta} \Omega^{\mu \gamma\delta}
\bar\beta\gamma_{\alpha\beta}
  \gamma_{\gamma\delta} \gamma \nonu
\A \A - {1 \over 4}\hat e \hat R \bar\beta \gamma \Bigr)
- i \, \hat e \, \bar B' \hat\slD B' \, \biggr],
\label{susyfp}
\ea
where we have used the condition (\ref{transvtraceless}) and
\be
V = ( \hat e_\alpha{}^\mu - \delta_\alpha{}^\mu )
\gamma^\alpha \partial_\mu + {1 \over 4} \hat e_\alpha{}^\mu
\hat \omega_{\mu \gamma\delta} \gamma^\alpha
\gamma^{\gamma\delta}.
\ee
\par
The action (\ref{susyrs}) shows that the superconformal mode $\eta$
gives the same contribution as that
of the spin ${1 \over 2}$ Majorana spinor $\lambda$
\be
\Gamma_{\eta} = {1 \over 2}\times I .
\label{etaloop}
\ee
\figthree
Diagrams with a $\phi_\alpha$ internal loop are constructed with
the vertices involved in $S_1$ in eq.\ (\ref{phivertex1}) and/or $S_2$
in eq.\ (\ref{phivertex2}).
First we consider vertices $S$ and $\Omega$ in $S_1$.
As shown in Fig.\ \ref{figurethree}, tadpole diagrams
with vertices from $S_1$ do not diverge.
The one-loop diagrams with two vertices from $S_1$ diverge
individually, but they cancel among them.
Therefore we find that the divergences due to diagrams with vertices
from $S_1$ cancel
\be
\Gamma_{S_1} = 0 .
\ee
\figfour
We next consider diagrams with vertices from $S_2$.
Since $S_2$ contains $\slpartial^{-1}$, it gives nonlocal vertices
involving $\slpartial^{-1}$.
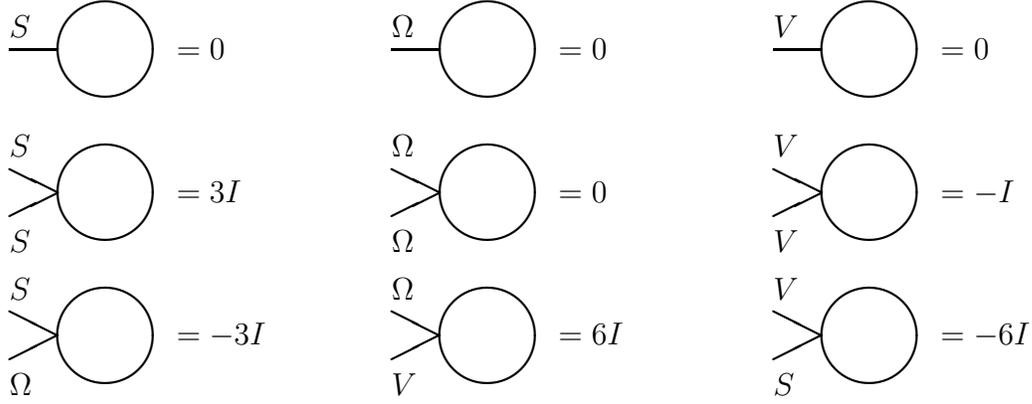
Tadpole diagrams with vertices from $S_2$ give
divergences as exhibited in Fig.\ \ref{figurefour}.
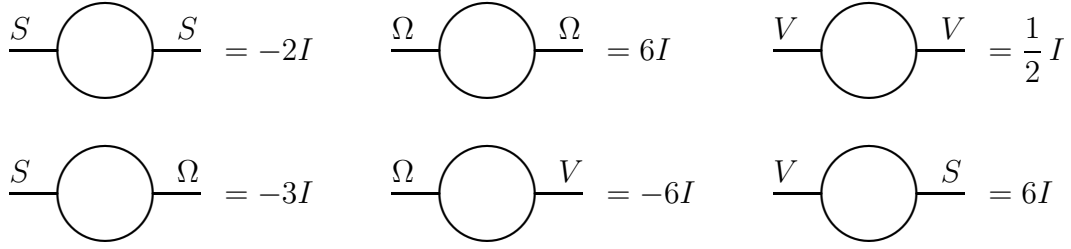
The  $\phi_\alpha$-loop diagrams with two vertices from the $S_2$
contribute divergences as shown in Fig.\ \ref{figurefive}.
Combining them together we find contributions of $\phi_\alpha$-loop
diagrams with vertices from $S_2$ as
\figfive
\be
\Gamma_{S_2} = {1 \over 2}\times I.
\ee
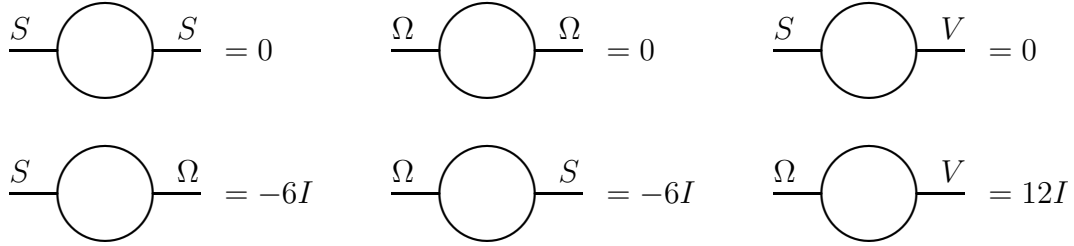
We also have $\phi_\alpha$-loop diagrams with one vertex from $S_1$
and the other from $S_2$, as shown in Fig.\ \ref{figuresix}.
Their divergences cancel each other
\figsix
\be
\Gamma_{S_1S_2} = 0 .
\ee
Therefore we find one-loop divergences from $\phi_\alpha$-loop diagrams
sum up to
\be
\Gamma_{\phi} = \Gamma_{S_1} + \Gamma_{S_2} + \Gamma_{S_1 S_2}
= {1 \over 2} \times I .
\label{philoop}
\ee
\par
\figseven
Let us now discuss the one-loop divergence of ghost loops
with vertices from $S_{{\rm SUSY-FP}}$ in eq.\ (\ref{susyfp}).
Tadpole diagrams with $S$, $\Omega$ or $\hat R$ vertices gives
divergences as shown in Fig.\ \ref{figureseven}.
The remaining tadpole diagram with the $\Omega\Omega$ vertex (the fourth
term in eq.\ (\ref{susyfp})) should be combined with
the one-loop diagram with two $\Omega$ vertices in order
to give a generally covariant result, as illustrated in
Fig.\ \ref{figureseven}.
There are also one-loop diagrams with two vertices from
$S_{{\rm SUSY-FP}}$.
Therefore contributions of loops of the supersymmetry ghosts
$\beta, \gamma$ sum up to
\be
\Gamma_{\beta\gamma} = 10 \times I .
\label{ghostloop}
\ee
Contribution from the Nakanishi-Lautrup field $B'$ is
the same as that of the Majorana spinor $\lambda$
\be
\Gamma_{\rm NL} = {1 \over 2} \times I .
\label{NLloop}
\ee
\par
Combining eqs.\ (\ref{etaloop}), (\ref{philoop}), (\ref{ghostloop})
and (\ref{NLloop}),
we obtain the one-loop divergence from the gravitino-ghost
system as
\be
\Gamma_{\rm gravitino}
= \Gamma_\eta + \Gamma_\phi + \Gamma_{\beta\gamma} + \Gamma_{\rm NL}
= {23 \over 2} \times I .
\label{gravitinoanomaly}
\ee
\par
Now we turn to discuss the one-loop divergence of graviton loops
in our method.
For the gauge parameters $\alpha=1, \beta=\epsilon$,
the action for graviton-ghost system contains the following quadratic
terms in the quantum fields $h_{\alpha\beta}, \phi$
\ba
S^{(2)}_{\rm E} + S^{(2)}_{\rm GC}
\A=\A \int d^d x
\left[ -{1 \over 4} \eta^{\mu\nu} \partial_\mu h_{\alpha\beta}
\partial_\nu h^{\alpha\beta}
+{\epsilon(\epsilon+2) \over 8}
 \eta^{\mu\nu} \partial_\mu \phi \partial_\nu \phi \right] \nonu
\A \A+ S_{\rm int} + S_{{\rm GC-FP}}, \label{grgh} \\[.5mm]
S_{\rm int}
\A = \A \int d^d x  \Biggl[ \,
 -{1 \over 4} S^{\mu\nu} \partial_\mu h_{\alpha\beta}
\partial_\nu h^{\alpha\beta}
+{\epsilon(\epsilon+2) \over 8}
 S^{\mu\nu} \partial_\mu \phi \partial_\nu \phi \nonu
\A \A - \Omega^{\mu\alpha}{}_{\gamma} \partial_\mu h_{\alpha\beta}
h^{\beta\gamma}
-{1 \over 2} T_{\alpha\beta\gamma\delta} h^{\alpha\beta}
h^{\gamma\delta}
+{1 \over 8} \epsilon^2 \hat R \phi^2
+{1 \over 2}\epsilon \,
\hat R^{\alpha\beta} h_{\alpha\beta} \phi \Biggr] \! ,
\\[.5mm]
S_{{\rm GC-FP}}
\A = \A \int d^d x \, i \, \hat e \,
b^\alpha \left( \hat D^\beta \hat D_\beta c_\alpha
+ \hat R _\alpha{}^\beta c_\beta \right) \nonu
\A = \A \int d^d x \, i \Biggl[ \,
-\eta^{\mu\nu} \partial_\mu b^\alpha \partial_\nu c_\alpha
- S^{\mu \nu} \partial_\mu b^\alpha \partial_\nu c_\alpha \nonu
\A  \A
- \Omega^{\mu\alpha\beta}(\partial_\mu b_\alpha c_\beta
- b_\alpha \partial_\mu c_\beta)
+ T_{\alpha\beta} b^\alpha c^\beta
+ {1 \over 2} \, \hat e B'^\alpha B'_\alpha \Biggr] \! ,
\label{gcfp}
\ea
%where $h=\eta^{\alpha\beta} h_{\alpha\beta}$ and
\ba
T_{\alpha\beta\gamma\delta}
\A=\A \hat e \left( \hat R_{\delta\alpha\beta\gamma}
+ \hat \omega_{\mu\alpha\epsilon}
\hat \omega^\mu{}_\gamma{}^\epsilon \eta_{\beta\delta}
+ \hat \omega_{\mu\alpha\gamma} \hat \omega^\mu{}_{\delta\beta} \right),
\nonu
T_{\alpha\beta}
\A=\A \hat e \left( \hat R_{\alpha\beta}
- \hat \omega_{\mu\alpha\gamma}
\hat \omega^\mu{}_\beta{}^\gamma \right), \quad
\Omega^{\mu\alpha\beta} = \hat e \hat\omega^{\mu\alpha\beta}.
\label{ttomega}
\ea
\par
\figeight
As shown in Fig.\ \ref{figureeight}, tadpole diagram with
the vertex $T$ should be
combined with the one-loop diagram with two vertices of
$\Omega$ to give a general coordinate invariant result.
Summing up the diagrams in Fig.\ \ref{figureeight},
we find the contributions from
the graviton ($h_{\alpha\beta}, \phi$) one-loop diagrams as
\be
\Gamma_{h \phi} = -9 \times I .
\ee
Similarly contributions from the ghosts $b^\alpha$ and
$c_\alpha$
for the general coordinate symmetry are given as shown in
Fig.\ \ref{figurenine} as
\be
\Gamma_{bc} = -16 \times I .
\ee
\fignine
There is no contribution from the auxiliary field $B'_\alpha$.
The graviton and the general coordinate ghosts add up to give
\be
\Gamma_{\rm graviton}
= \Gamma_{h\phi} + \Gamma_{bc} = -25 \times I .
\label{gravitonanomaly}
\ee
This result agrees with that in ref.\ \cite{KN}.
By combining eqs.\ (\ref{gravitinoanomaly}) and (\ref{gravitonanomaly})
with the supersymmetric matter multiplets
($X^i$, $\lambda^i$) ($i = 1, \cdots, \hat c$)
in eqs.\ (\ref{scalaranomaly}) and (\ref{spinoranomaly}),
we find the total one-loop divergence as
\be
\Gamma_{\rm total} = - {3 \over 2} (9 - \hat c) \, I.
\label{oneloopdivergence}
\ee
This result may be expected from the conformal anomaly for the
supergravity coupled to the supersymmetric matter multiplets
in two dimensions.
%
%%%%%%%  Section 5  %%%%%%%%%%%%%%%%%%%%%%%%%%%%%%%%%%%%%%%%
%
\newsection{Fixed point and two-dimensional limit}
In order to define a renormalized gravitational coupling constant $G$,
we need to specify a reference scale for the metric
$g_{\mu\nu}$ \cite{KN}.
The choice of the reference scale affects the definition of the
renormalized coupling constant and consequently the associated
beta function in general.
In order to obtain a scale invariant result in the limit of
two dimensions, it is most appropriate to choose the reference
scale as the coefficient for a spinless operator $\Psi$
which has the canonical dimension $2\Delta_0=1$ in two dimensions
\cite{KN}.
It has been found that such a reference operator does not suffer
from divergences at one-loop order in the parametrization
(\ref{conformalcond}) \cite{KKN}.
This fact allows us to renormalize the gravitational coupling constant
alone without considering the renormalization of the
reference operator $\Psi$, since it is automatically renormalized
in our parametrization at least at one-loop order.
Therefore the bare coupling constant $G_0$ is related to the
renormalized coupling constant $G$ through
\be
{1 \over 16\pi G_{0}} = {\mu^\epsilon \over 16\pi}
\left( {1 \over G} - {9 - \hat c \over \epsilon} \right).
\ee
The second term in the right hand side is the one-loop counter term
which cancels the one-loop divergence (\ref{oneloopdivergence}).
\par
The beta function defined by
$\beta(G)=\mu {\partial \over \partial \mu} G$
is determined by
$\mu {\partial \over \partial \mu} G_{0} = 0$
\be
\beta(G) = \epsilon G - (9 - \hat c) G^2 .
\ee
We see that quantum supergravity in $2+\epsilon$ dimensions exhibits
a nontrivial ultraviolet fixed point $G^*$ at
\be
G^* = {\epsilon  \over 9 - \hat c}.
\ee
\par
Let us now turn to consider renormalizing the physical operators
which are obtained in sect.\ 2.
We are especially interested in taking the two-dimensional limit.
Following ref.\ \cite{KKN}, the two-dimensional theory can be
obtained as a limit $\epsilon \rightarrow 0$
of the $2+\epsilon$ gravity in
a strong coupling region $G \gg \epsilon$.
In that limit, one can obtain the anomalous dimension in powers of
$1/(25-c)$. Moreover, by assuming the dominance of conformal mode,
one can derive a result which exactly reproduces the conformal field
theory approach \cite{DDK}.
We can apply their reasoning to our supergravity case as well.
In this paper, we restrict ourselves to operators $\Psi_{NS}$
in the Neveu-Schwarz sector.
\par
Although we have obtained the bare physical operators
(\ref{physicalop}), which are
invariant under all the local gauge symmetries classically,
we need to find an extra dependence on the supergravity
multiplet induced by quantum effects of the matter fields.
As an example let us consider a physical operator for the
matter operator $\e^{i p \cdot X}$ in the bosonic theory.
Classically, the integrand of the physical operator is
$e \e^{i p \cdot X} = \hat e \e^{-{d \over 2}\phi} \e^{i p \cdot X}$.
Quantum effects of the matter field $X$ changes the conformal
mode dependence to $\e^{-{d \over 2}(1-\Delta_0)\phi}$, where
$2 \Delta_0 = p^2$ is the conformal dimension of the
matter operator \cite{WEINBERG}.
\par
To find out the gravitational dressing in the supersymmetric case
we use the superfield formalism.
Superfields $X(x, \theta)$ for a matter supermultiplet and
$\Phi(x, \theta)$ for the (super)conformal modes can be written
in terms of a Grassmann number spinor $\theta$ \cite{HBG}
\ba
X(x, \theta) \A = \A X(x) - i \bar\theta \lambda(x)
- {1 \over 2} i \bar\theta \theta F(x), \nonu
\Phi(x, \theta) \A = \A \phi(x) - i \bar\theta \eta(x)
+ {1 \over d} i \bar\theta \theta S'(x).
\label{superfield}
\ea
The auxiliary field $S'$ is related to the auxiliary field $S$ in
eq.\ (\ref{supergravityaction}) as
\ba
S' \A = \A S + {1 \over 8 \kappa_0} i \bar\psi_\mu \gamma^\nu
           \gamma^\mu \psi_\nu \nonu
   \A = \A  S + {1 \over 4} i \kappa_0 \bar\phi_\alpha \phi^\alpha
           + {1 \over 8} i d (d-2) \kappa_0 \bar\eta \eta.
\ea
As in ref.\ \cite{KKN} we use the supergravity action with the bare
gravitational constant $G_0$ to compute the anomalous dimension
of the operators. The fields $\phi$, $\phi_\alpha$ and $\eta$
in eq.\ (\ref{superfield}) are defined as in
eqs.\ (\ref{conformalcond}) and (\ref{superconformal})
with $\kappa$ replaced by $\kappa_0 = \sqrt{16 \pi G_0}$.
The field $S$ has also been rescaled by $\kappa_0$.
The physical operator in eq.\ (\ref{physicalop})
is given in superspace
\be
O_p = \int d^d x d^2 \theta \, \hat e
\e^{-{d \over 2} \kappa_0 \Phi(x, \theta)}
\e^{i p \cdot X(x, \theta)},
\label{superspacevertexop}
\ee
We now quantize the matter fileds and find the conformal
dimension $2\Delta_0$ for the matter part of the operator.
In the case of the momentum eigenstate (\ref{superspacevertexop}),
we find that $\Delta_0 = {1 \over 2} p^2$.
For the spinless operator, it is enough to consider the dressed
operator by multiplying the appropriate factor of the exponential
of the superfield $\Phi(x, \theta)$ for the conformal mode in
eq.\ (\ref{superfield}) \cite{KKN}, \cite{DDK}, \cite{DHK}
\ba
O^{\rm dressed}_p (x)
\A = \A \int d^2 \theta \, \hat e
\e^{-{d \over 2}(1-\Delta'_0) \kappa_0 \Phi(x, \theta)}
\e^{i p \cdot X(x, \theta)} \nonu
\A = \A \hat e
\e^{-{d \over 2}(1-\Delta'_0) \kappa_0 \phi}
\biggl[ i p \cdot \bar\lambda \lambda \cdot p - 2 i p \cdot F
-2(1-\Delta'_0) \kappa_0 S' \nonu
\A \A - {d^2 \over 4} i (1-\Delta'_0)^2 \kappa_0^2 \bar\eta \eta
- d (1-\Delta'_0) \kappa_0 \bar\eta \lambda \cdot p \biggr]
\e^{i p \cdot X},
\label{dressedop}
\ea
where
\be
\Delta'_0 = {1 \over 2} + \Delta_0, \qquad
\Delta_0 = {1 \over 2} p^2.
\ee
The conformal dimension $\Delta'_0$ of the operator $O_p(x)$
is ${1 \over 2}$ larger than the conformal dimension
$\Delta_0$ for the operator in superspace because of
the $\theta$ integration.
\par
To compute one-loop renormalization due to quantum effects of
the supergravity multiplet, we give nonzero background values
to the matter fields
$\hat X \not= 0,\ \hat \lambda \not= 0,\ \hat F \not= 0$.
Supergravity background fields are
$\hat e_\mu{}^\alpha \not= \delta_\mu^\alpha,\ \hat\psi_\mu = 0,\
\hat S = -{1 \over 8} i \kappa_0 \bar{\hat\lambda} \hat\lambda$.
The background of $S$ has been chosen such that it satisfies
the equation of motion.
In this background the dressed operator (\ref{dressedop}) becomes
\be
\hat O^{\rm dressed}_p (x)
= \hat e \left[
i p \cdot \bar{\hat\lambda} \hat\lambda \cdot p
- 2 i p \cdot \hat F + {1 \over 4} i (1-\Delta'_0)
\kappa_0^2 \bar{\hat\lambda} \hat\lambda
\right] \e^{i p \cdot \hat X}.
\ee
We have to compute quantum effects of the supergravity multiplet
by introducing their fluctuations $h_{\alpha\beta},\ \phi,\
\phi_{\alpha},\ \eta$.
We only need to compute $O(\epsilon^{-2})$ singularities
since $\kappa_0^2 = O(\epsilon)$ as we will see below.
At one-loop order the expectation value of the dressed
operator (\ref{dressedop}) becomes
\be
\VEV{O^{\rm dressed}_p(x)}
 =  \left[ 1 + {(1-\Delta'_0)^2 \kappa_0^2 \over 2\pi\epsilon^2}
+ O(\epsilon^{-1}) \right]
\hat O^{\rm dressed}_p (x),
\ee
\par
The divergence can be removed by defining a renormalized operator
\be
O^{\rm ren}_p(x) = Z_{\Delta_0} O^{\rm dressed}_p(x), \qquad
Z_{\Delta_0} = 1 -
{(1-\Delta'_0)^2 \kappa_0^2 \mu^\epsilon \over 2\pi \epsilon^2}.
\ee
Consequently the anomalous dimension is given by
\be
\gamma_{\Delta_0}
= \mu {\partial \over \partial \mu} \ln Z_{\Delta_0}
= - 8 (1-\Delta'_0)^2 {G_0 \mu^\epsilon \over \epsilon}.
\ee
If we consider the strong coupling region $G \gg \epsilon$,
\be
{1 \over G_0 \mu^\epsilon}
= {1 \over G} - {9-\hat c \over \epsilon}
\approx - {9-\hat c \over \epsilon},
\ee
we obtain the anomalous dimension to the first order in $1/(9-\hat c)^2$
\be
\gamma_{\Delta_0} = {4 (1-\Delta'_0)^2 \over Q^2}
= {(1 - 2 \Delta_0)^2 \over Q^2},
\label{anomalousdim}
\ee
where $Q$ is the source charge for the Liouville field in the
case of superconformal field theory \cite{DHK}
\be
Q = \sqrt{9-\hat c \over 2}.
\ee
\par
In the conformal field theory approach \cite{DHK},
the gravitational dressing of the operator
with the conformal dimension $(\Delta_0, \Delta_0)$,
$2\Delta_0 = {1 \over 2} p^2$ for the matter part
\be
\int d^2 x d^2 \theta \hat E \e^{\beta \Phi}
\e^{i p \cdot X(x,\theta)}
\ee
is determined by requiring that they must have conformal dimension
$({1 \over 2}, {1 \over 2})$ \cite{DHK}
\be
\Delta_0-{1 \over 2}\beta(\beta+Q)={1 \over 2}.
\ee
The choice of the two solution can be made by resorting to the
classical limit \cite{DDK}
\be
\beta = -{Q \over 2}
    \left( 1 - \sqrt{1 - {{4(1 - 2 \Delta_0)} \over Q^2}} \right).
\ee
\par
We can compare our result with the conformal field theory approach
by using the scaling argument \cite{KKN}.
Let us call the dressing exponent $\beta$ for the operator with
the conformal dimension $(0, 0)$ as $\alpha$.
The scaling exponent of the operator insertion $O_p$ is given by
the ratio between the dressing exponent $\beta$ and $\alpha$
\ba
{\beta \over \alpha} \A=\A
\frac{ {1 \over 2} - \Delta_0
+ {1 \over 2} \left({1 \over 2} - \Delta_0\right)^2 {4 \over Q^2}
+ O\left(\left({4 \over Q^2}\right)^2\right) }
{ {1 \over 2}
+ {1 \over 8} {4 \over Q^2}
+ O\left(\left({4 \over Q^2}\right)^2\right) }.
\ea
The scaling exponent can be given in terms of the anomalous dimension
$\gamma_{\Delta_0}$ as \cite{KKN}
\be
{\beta \over \alpha} = {2 \left( {1 \over 2}-\Delta_0 \right)
   + \gamma_{\Delta_0} \over 1 + \gamma_{\Delta_0 = 0}}.
\ee
Our result in eq.\ (\ref{anomalousdim}) is nothing but the first
nontrivial term of the expansion in powers of $1/Q^2$.
Moreover, we can show that the result of the conformal field theory
approach can be reproduced to all orders of $1/Q^2$
in a way precisely analogous to ref.\ \cite{KKN}.
In renormalizing the operators, we see that the fields other than the
conformal mode $\phi$ does not play a role.
Let us assume that the divergences at higher orders are also dominated
by the one-loop counter term in the bare lagrangian and consider only
the conformal mode.
Then the divergent part can be recast into a zero-dimensional
path-integral. We find that the argument in
ref.\ \cite{KKN} is valid in our case with the replacement of
$1-\Delta_0$ by $1-\Delta'_0$ and $Q^2=(25-c)/3$ by $Q^2=(9-\hat c)/2$.
Therefore we see that our result is fully consistent with the conformal
field theory approach.
%
%%%%%%%  Section 6   %%%%%%%%%%%%%%%%%%%%%%%%%%%%%%%%%%%%%%%%%%
%
\newsection{A method of dimensional reduction}
So far we have been discussing $d = 2+\epsilon$ dimensional
supergravity using the action (\ref{supergravityaction})
interpolating between two- and three-dimensional ones.
A problem of this approach is that the action is invariant
under the local supersymmetry transformations only up to
terms of order $(\mbox{fermi fields})^3$ in general $d$ dimensions.
This noninvariance is due to the fact that
the bosonic and fermionic fields have different numbers of
components in general $d$ dimensions
while supersymmetry requires the same number.
Since these noninvariant terms can only affect higher loop orders,
our computations of divergences sould be valid at one loop level.
Neverthless, it is desirable to have a manifestly supersymmetric
regularization method in noninteger dimensions.
\par
Another way to consider supergravity in $2+\epsilon$ dimensions
is to use a method of dimensional reduction \cite{DIMRED}.
This method was
proposed by Siegel \cite{SIEGELDR} as a regularization which
preserves rigid supersymmetry for four-dimensional theories.
In this regularization one starts from a supersymmetric theory in
$D = 4$ dimensions and supposes that fields depend on only
$4-\epsilon$ coordinates
while keeping the number of components of the fields unchanged.
In contrast to the ordinary dimensional regularization \cite{THVE},
one may hope that supersymmetry is preserved
in the resulting $d = 4-\epsilon$ dimensional theory since the
number of components of bosonic and fermionic fields remain the same.
\par
We can use this idea of dimensional reduction to construct
a theory in $d = 2 + \epsilon$ dimensions
starting from the $D = 2$ or $D = 3$ dimensional supergravity.
We denote the $D$-dimensional fields as $E_M{}^A$,
$\Psi_M \; ( = E_M{}^A \Psi_A )$, $\hat S$,
where $M, N, \cdots = 0, 1, \cdots, D-1$ and
$A, B, \cdots = 0, 1, \cdots, D-1$ are $D$-dimensional world indices
and local Lorentz indices respectively.
The spinor fields are two-component ones.
To distinguish quantities in $D$ dimensions and those in $d$ dimensions,
we put a hat on $D$-dimensional ones, if necessary.
The action of these fields has the form in
eq.\ (\ref{supergravityaction}) and is shown to be
invariant under the supertransformations (\ref{supertrans})
using the $D$-dimensional gamma matrix identities.
We split the indices as $M = (\mu, m)$ and $A = (\alpha, a)$,
where $\mu, \alpha = 0, \cdots, d-1$ and $m, a = d, \cdots, D-1$.
To reduce the $D$-dimensional theory to $d = 2 + \epsilon$ we
suppose that all the fields are independent of $D-d$ coordinates $x^m$
and parametrize them as \cite{DIMRED}
\ba
E_M{}^A \A = \A \Delta^{-{1 \over 2(D-2)}} \left(
\begin{array}{ccc}
e_\mu{}^\alpha & A_\mu{}^m e_m{}^a \\
0         & e_m{}^a
\end{array}
\right), \nonu
\Psi_A \A = \A \Delta^{1 \over 4(D-2)} \left(
\psi_A - {1 \over D-2} \gamma_A \gamma^b \psi_b \right), \nonu
\hat S \A = \A \Delta^{1 \over 2(D-2)} S, \qquad
\Delta \equiv ( \det\; e_m{}^a )^2.
\label{six1}
\ea
We have used a part of the $D$-dimensional local Lorentz symmetry
to put $E_m{}^\alpha = 0$. We have also made
(super) Weyl rescalings to obtain the standard kinetic terms
in the $d$-dimensional action.
The field content of the $d$-dimensional theory
is a vielbein $e_\mu{}^\alpha$, $D-d$ vector fields $A_\mu{}^m$,
$(D-d)^2$ scalar fields $e_m{}^a$, a Rarita-Schwinger field
$\psi_\mu = e_\mu{}^\alpha \psi_\alpha$, $D-d$ spin $1/2$ spinor
fields $\psi_a$ and a scalar field $S$.
Because of the local symmetry SO($D-d$), which is a part of the
$D$-dimensional local Lorentz symmetry, not all
components of $e_m{}^a$ are physical. There are only
$(D-d)(D-d+1)/2$ physical scalar fields in $e_m{}^a$.
The spinor fields $\psi_\mu$ and $\psi_a$ remain two-component.
The action and the supertransformations in $d$ dimensions
can be obtained from eqs.\ (\ref{supergravityaction}) and
(\ref{supertrans}).
We give the results for them in appendix B.
One can also introduce matter fields in $d$ dimensions by
dimensionally reducing $D$-dimensional matter multiplets
(\ref{matteraction}).
\par
The (super) Weyl rescalings in eq.\ (\ref{six1}) is singular
at $D=2$. This is a reflection of the fact that the Weyl
rescaling in $D$ dimensions cannot absorb the spacetime
dependent factor multiplying the Einstein term.
This singular behavior can be avoided if we start
from $D=3$, for instance.
However, we still have a singularity at $d=2$.
For $D = 3$, $d = 2$, eq.\ (\ref{six1}) gives
$E_m{}^a= 1,\ \Psi_m = 0$ and these degrees of freedom are not
represented by the $d$-dimensional fields.
Therefore, it is better to use a parametrization without
(super) Weyl rescalings for $d=2$ case.
The resulting two-dimensional theory turns out to be a
supersymmetric version of so-called dilaton gravity \cite{CHAM}.
We present the action and the supertransformations in
this case in appendix C.
\par
It was later pointed out that the regularization by dimensional
reduction is mathematically inconsistent \cite{SIEGELIC}.
The inconsistency discussed in ref.\ \cite{SIEGELIC} uses the
antisymmetric epsilon tensor $\epsilon^{\mu\nu\rho\sigma}$.
One might think that this is due to a difficulty to define
chiral quantities such as the epsilon tensor and $\gamma_5$
in general non-integer dimensions. Such a difficulty is well
known in the ordinary dimensional regularization \cite{THVE}.
Chiral quantities are essential for four-dimensional theories
since the definition of chiral scalar supermultiplets uses $\gamma_5$.
Therefore one might hope that the regularization by dimensional
reduction could be consistent in vector like theories, such as
supergravity theories we are considering, which do not use chiral
quantities. Unfortunately, this is not the case as we will show below.
\par
Let us consider a reduction from $D = 3$ to $d$ dimensions.
According to the definition of this regularization the
gamma matrices $\gamma^A = (\gamma^\alpha, \gamma^a)$
are $2 \times 2$ matrices and satisfy
\ba
\{ \gamma^A, \gamma^B \} \A = \A 2 \eta^{AB},
\label{six2} \\
\gamma^{A_1 \cdots A_n} \A = \A 0 \quad (n > 3).
\label{six3}
\ea
These equations are also required to prove supersymmetry of
the action. Let us assume
\be
\eta_{\alpha\beta} \eta^{\beta\gamma} = \delta_\alpha^\gamma, \qquad
\delta_\alpha^\alpha = d,
\label{six4}
\ee
and define $\gamma_\alpha \equiv \eta_{\alpha\beta} \gamma^\beta$.
Since $\gamma^{\alpha\beta} \gamma^{cd} = \gamma^{\alpha\beta cd}$,
from eq.\ (\ref{six3}) we have
\be
\gamma^{\alpha\beta} \gamma^{cd} = 0.
\label{six5}
\ee
Multiplying this equation by $\gamma_{\alpha\beta}$ and $\gamma_{cd}$,
and using eqs.\ (\ref{six2}) and (\ref{six4}), we obtain
\ba
0 \A = \A \gamma_{\alpha\beta} \gamma^{\alpha\beta}
          \gamma^{cd} \gamma_{cd} \nonu
  \A = \A d (d-1) (d-2) (d-3).
\label{six6}
\ea
Therefore this regularization can be consistent only for
integer dimensions $d = 0, 1, 2, 3$.
\par
The essential point of this discussion is eq.\ (\ref{six3}).
To emphasize this point let us consider another difficulty.
{}From eqs.\ (\ref{six2}), (\ref{six4}) we can show that
\be
\gamma_\alpha \gamma^{\alpha\beta_1 \cdots \beta_n}
= (d-n) \, \gamma^{\beta_1 \cdots \beta_n}.
\label{six7}
\ee
When $d$ is not an integer,
using eqs.\ (\ref{six3}), (\ref{six7}) we find that
$\gamma^{\alpha_1 \cdots \alpha_n} = 0$ for all non-negative
integers $n$. In particular, we have $\gamma^\alpha = 0$.
\par
In the above discussions we have not used the antisymmetric
epsilon tensor or $\gamma_5$. The above difficulties do not
arise in the ordinary dimensional regularization \cite{THVE}
since eq.\ (\ref{six3}) need not be satisfied.
If one drops the requirement (\ref{six3}), however, supersymmetry
of the $(2+\epsilon)$-dimensional theory will not be guaranteed.
To construct supergravity in $2+\epsilon$ dimensions
in this approach, one has to find out a modification of
the regularization by dimensional reduction
such that it is consistent and preserves supersymmetry.
\par
\vspace{5mm}
%
%%%%%%%  Acknowledgement  %%%%%%%%%%%%%%%%%%%%%%%%%%%%%%%%%%%
%
The authors would like to thank K.S. Stelle for reading the
manuscript and for a useful comment.
One of the authors (N.S.) thanks T. Uematsu for a useful discussion
on the tensor calculus of supergravity in two and three dimensions.
He is also grateful to M. Ninomiya, H. Kawai, and Y. Kitazawa
for a discussion on gravity theories in $2+\epsilon$ dimensions.
One of the authors (Y.T.) would like to thank the Theoretical
Physics Group of Imperial
College for hospitality, and the Japan Society for the Promotion
of Science and the Royal Society for a grant.
This work is supported in part by Grant-in-Aid for Scientific
Research (S.K.) and (No.05640334) (N.S.), and Grant-in-Aid for
Scientific Research for Priority Areas (No.05230019) (N.S.) from
the Ministry of Education, Science and Culture.
\par
%
%%%%%%%  Appendix A  %%%%%%%%%%%%%%%%%%%%%%%%%%%%%%%%%%%%%%%%%%
%
\def\numberbysectiona{\@addtoreset{equation}{section}
\def\theequation{A.\arabic{equation}}}
\numberbysectiona
\vspace{7mm}
\pagebreak[3]
\setcounter{section}{1}
\setcounter{equation}{0}
\setcounter{subsection}{0}
\setcounter{footnote}{0}
\begin{center}
{\large{\bf Appendix A. One-loop divergences of a vector field}}
\end{center}
\nopagebreak
\medskip
\nopagebreak
\hspace{3mm}
In this appendix we compute one-loop divergences of a vector
gauge field $A_\mu$ in a background gravitational field
$\hat g_{\mu\nu}$. The gauge invariant action is
\ba
S_{\rm V}
\A = \A - {1 \over 4} \int d^d x \, \hat e \hat g^{\mu\rho}
        \hat g^{\nu\sigma} F_{\mu\nu} F_{\rho\sigma}, \nonu
\A = \A - {1 \over 2} \int d^d x \, \hat e \left[ \hat g^{\mu\nu}
          \hat D_\rho A_\mu \hat D^\rho A_\nu
        - (\hat D^\mu A_\mu)^2 + \hat R^{\mu\nu} A_\mu A_\nu \right].
\label{a1}
\ea
To fix the gauge symmetry we introduce Faddeev-Popov
(anti-)ghost fields $b, c$ and a Nakanishi-Lautrup auxiliary
field $B$. Their BRST transformations are
\be
\delta_{\rm B} A_\mu = \partial_\mu c, \quad
\delta_{\rm B} c = 0, \quad
\delta_{\rm B} b = iB, \quad
\delta_{\rm B} B = 0.
\label{a2}
\ee
We use a gauge function
\be
F_{\rm YM} = \hat e \left( \hat D^\mu A_\mu + {1 \over 2} \xi B \right),
\label{a3}
\ee
where $\xi$ is a constant gauge parameter.
Then, the gauge fixing term and the ghost action are
\ba
S_{\rm YM}
\A = \A \int d^d x \, \delta_{\rm B} \left( - i b F_{\rm YM} \right) \nonu
\A = \A \int d^d x \, \hat e \left[ {1 \over 2} \xi B'^2
- {1 \over 2\xi} (\hat D^\mu A_\mu)^2
+ i b \hat D^\mu \partial_\mu c \right],
\label{a4}
\ea
where $B'$ is a shifted auxiliary field.
The total gauge fixed action is a sum of eqs.\ (\ref{a1}) and
(\ref{a4}): $S_{\rm tot} = S_{\rm V} + S_{\rm YM}$.
We choose the gauge parameter as $\xi = 1$, which simplifies
loop calculations.
\par
To compute one-loop counterterms we expand the background
gravitational field as in eq.\ (\ref{backgroundexp}).
The total action is decomposed as
\ba
S_{\rm tot} \A = \A - {1 \over 2} \int d^d x \biggl[ \,
\eta^{\mu\nu} \partial_\mu A_\alpha \partial_\nu A^\alpha
+ S^{\mu\nu} \partial_\mu A_\alpha \partial_\nu A^\alpha
+ 2 \Omega^{\mu\alpha\beta} A_\beta \partial_\mu A_\alpha \nonu
\A \A + T'^{\, \alpha\beta} A_\alpha A_\beta
- 2 i \eta^{\mu\nu} \partial_\mu b \partial_\nu c
- 2 i S^{\mu\nu} \partial_\mu b \partial_\nu c
+ {1 \over 2} \, \hat e B'^2 \, \biggr],
\label{a5}
\ea
where we have used a field with a local Lorentz index
$A_\alpha = \hat e_\alpha{}^\mu A_\mu$.
The functions $S^{\mu\nu}$, $\Omega^{\mu\alpha\beta}$ are
defined in eqs.\ (\ref{svertex}), (\ref{ttomega}) and
\be
T'^{\, \alpha\beta}
= \hat e \left( \hat R^{\alpha\beta}
+ \hat\omega_{\mu\gamma}{}^\alpha
\hat\omega^{\mu\gamma\beta} \right).
\label{a6}
\ee
\figten
Divergent $A_\alpha$ one-loop diagrams are shown in
Fig.\ \ref{figureten}. The tadpole
diagram with the vertex $T'$ should be combined with
the diagram with two $\Omega$ vertices to give a general coordinate
invariant result.
The contribution from the ghost loop diagrams is the same as
that of a complex scalar field of fermionic statistics and is
given by $-2I$. Therefore the total one-loop divergence of
a vector gauge field is
\be
\Gamma_{\rm vector} = - 6 \times I.
\label{a7}
\ee
This result is consistent with eq.\ (16.75) in ref.\ \cite{WEI}.
%
%%%%%%%  Appendix B  %%%%%%%%%%%%%%%%%%%%%%%%%%%%%%%%%%%%%%%%%%
%
\def\numberbysectionb{\@addtoreset{equation}{section}
\def\theequation{B.\arabic{equation}}}
\numberbysectionb
\vspace{7mm}
\pagebreak[3]
\setcounter{section}{1}
\setcounter{equation}{0}
\setcounter{subsection}{0}
\setcounter{footnote}{0}
\begin{center}
{\large{\bf Appendix B. Dimensional reduction to $d$ dimensions}}
\end{center}
\nopagebreak
\medskip
\nopagebreak
\hspace{3mm}
In this appendix we work out a dimensional reduction of
supergravity in $D$ dimensions to $d$ dimensions.
The $D$-dimensional fields are taken to be independent of $D-d$
coordinates $x^m$ and are
parametrized by $d$-dimensional fields as in eq.\ (\ref{six1}).
We obtain the $d$-dimensional action up to four-fermi terms
and the supertransformations to this order.
\par
It is convenient to describe the $d$-dimensional scalar fields
$e_m{}^a$ as a G/H nonlinear $\sigma$-model, where
G = GL($D-d$) and H = SO($D-d$).
The scalar fields $e_m{}^a(x) \in {\rm G}$ transform under
${\rm G}_{\rm rigid} \times {\rm H}_{\rm local}$ as
\be
e_m{}^a(x) \longrightarrow L_m{}^n e_n{}^b(x) O_b{}^a(x), \quad
L \in {\rm G},\ O(x) \in {\rm H}.
\label{b1}
\ee
We decompose a derivative of the scalar fields into two parts:
\ba
e_a{}^m \partial_\mu e_{mb} \A = \A P_{\mu ab} + Q_{\mu ab}, \nonu
P_{\mu ab} \A \equiv \A {1 \over 2} ( e_a{}^m \partial_\mu e_{mb} +
e_b{}^m \partial_\mu e_{ma} ) = P_{\mu ba}, \nonu
Q_{\mu ab} \A \equiv \A {1 \over 2} ( e_a{}^m \partial_\mu e_{mb} -
e_b{}^m \partial_\mu e_{ma} ) = - Q_{\mu ba}.
\label{b2}
\ea
$Q_{\mu ab}$ is in the Lie algebra of H, while $P_{\mu ab}$
is in the orthogonal complement of H in G.
They are invariant under ${\rm G}_{\rm rigid}$.
$Q_{\mu ab}$ transforms as an H gauge field under ${\rm H}_{\rm local}$
and can be used to define covariant derivatives.
$P_{\mu ab}$ can be expressed as a covariant derivative of $e_m{}^a$:
\be
P_{\mu ab} = e_a{}^m \left( \partial_\mu e_{mb}
+ Q_{\mu b}{}^c e_{mc} \right) \equiv e_a{}^m D_\mu e_{mb}
\label{b3}
\ee
and transforms covariantly under ${\rm H}_{\rm local}$.
The kinetic terms of the scalar fields will be written by using
$P_{\mu ab}$.
The torsionless spin connection in $D$ dimensions $\hat\omega_{ABC}
= E_A{}^M \hat\omega_{MBC}(E)$ becomes
\ba
\hat\omega_{\alpha\beta\gamma} \A = \A \Delta^{1 \over 2(D-2)} \left[
\omega_{\alpha\beta\gamma} - {1 \over D-2}(\eta_{\alpha\beta}
e_\gamma{}^\mu - \eta_{\alpha\gamma} e_\beta{}^\mu) P_{\mu d}{}^d
\right], \nonu
\hat\omega_{\alpha\beta c} \A = \A {1 \over 2} \Delta^{1 \over 2(D-2)}
F_{\alpha\beta}^m e_{mc}, \nonu
\hat\omega_{\alpha bc} \A = \A \Delta^{1 \over 2(D-2)}
Q_{\alpha bc}, \nonu
\hat\omega_{a\beta\gamma} \A = \A - {1 \over 2} \Delta^{1 \over 2(D-2)}
F_{\beta\gamma}^m e_{ma}, \nonu
\hat\omega_{a\beta c} \A = \A \Delta^{1 \over 2(D-2)} \left[
- P_{\beta ac} + {1 \over D-2} \delta_{ac} P_{\beta d}{}^d
\right], \nonu
\hat\omega_{abc} \A = \A 0,
\label{b4}
\ea
where $\omega_{\alpha\beta\gamma}
= e_\alpha{}^\mu \omega_{\mu\beta\gamma}(e)$ is the torsionless
spin connection in $d$ dimensions defined by $e_\mu{}^\alpha$.
\par
Using the above formulae the $D$-dimensional action of the form
(\ref{supergravityaction}) becomes
\ba
S_{\rm SG} \A = \A {1 \over 16\pi G_0} \int d^d x \, e \biggr[
R - {1 \over 4} g_{mn} F^m_{\mu\nu} F^{n\mu\nu}
- P_{\mu ab} P^{\mu ab} + {1 \over D-2} ( P_{\mu a}{}^a )^2 \nonu
\A \A + i \bar\psi_\alpha \gamma^{\alpha\beta\gamma} D_\beta \psi_\gamma
+ i \bar\psi_a \gamma^\beta D_\beta \psi^a
+ {1 \over D-2} i \bar\psi_a \gamma^a \gamma^\beta \gamma^c
D_\beta \psi_c
+ {1 \over 8} i F_{\alpha\beta}^m e_m{}^a \nonu
\A \A \times \left(
\bar\psi^\gamma \gamma_a \gamma_{[\delta} \gamma^{\alpha\beta}
\gamma_{\gamma]} \psi^\delta
+ 2 \bar\psi_\gamma \gamma^{\alpha\beta} \gamma^\gamma \psi_a
+ \bar\psi_b \gamma_a \gamma^{\alpha\beta} \psi^b
+ {1 \over D-2} \bar\psi_b \gamma^b \gamma_a \gamma^{\alpha\beta}
\gamma^c \psi_c \right) \nonu
\A \A - i P_{\alpha bc} \bar\psi_\beta \gamma^\alpha \gamma^\beta
\gamma^b \psi^c
+ {1 \over D-2} i P_{\alpha b}{}^b \bar\psi_\beta \gamma^\alpha
\gamma^\beta \gamma^c \psi_c
- {D-2 \over D-1} \; S^2 + O(\psi^4) \biggr],
\label{b5}
\ea
where $g_{mn} = e_m{}^a e_n{}^b \delta_{ab}$.
The covariant derivatives on the spinor fields are given by
\ba
D_\mu \psi_\alpha \A = \A \left( \partial_\mu
+ {1 \over 4} \omega_\mu{}^{\beta\gamma} \gamma_{\beta\gamma}
+ {1 \over 4} Q_\mu{}^{bc} \gamma_{bc} \right) \psi_\alpha
+ \omega_{\mu\alpha}{}^\beta \psi_\beta, \nonu
D_\mu \psi_a \A = \A \left( \partial_\mu
+ {1 \over 4} \omega_\mu{}^{\beta\gamma} \gamma_{\beta\gamma}
+ {1 \over 4} Q_\mu{}^{bc} \gamma_{bc} \right) \psi_a
+ Q_{\mu a}{}^b \psi_b.
\label{b6}
\ea
The coefficient of each term of the action is $d$-independent.
\par
Supertransformations in $d$ dimensions is defined as a sum of
$D$-dimensional supertransformations (\ref{supertrans}) and particular
local Lorentz transformations:
\ba
\delta_Q(\varepsilon)
\A = \A \delta_{\hat Q}(\eta)
+ \delta_{\hat L}(\lambda(\varepsilon)), \quad
\eta = \Delta^{-{1 \over 4(D-2)}} \varepsilon, \nonu
\lambda_{\alpha\beta}(\varepsilon) \A = \A - {1 \over D-2} i
\bar\varepsilon \gamma_{\alpha\beta} \gamma^c \psi_c, \nonu
\lambda_{\alpha b}(\varepsilon) \A = \A
- \lambda_{b\alpha}(\varepsilon) =
i \bar\varepsilon \gamma_\alpha \left( \psi_b
- {1 \over D-2} \gamma_b \gamma^c \psi_c \right), \nonu
\lambda_{ab}(\varepsilon) \A = \A
- {1 \over D-2} i \bar\varepsilon \gamma_{ab} \gamma^c \psi_c.
\label{b7}
\ea
The parameter $\lambda_{\alpha b}$ has been chosen to preserve
the condition $E_m{}^\alpha = 0$, while $\lambda_{\alpha\beta},
\lambda_{ab}$ have been chosen to simplify transformations of
$e_\mu{}^\alpha, e_m{}^a$ respectively.
We obtain the dimensionally reduced supertransformations as
\ba
\delta_Q e_\mu{}^\alpha
\A = \A - i \bar\varepsilon \gamma^\alpha \psi_\mu, \nonu
\delta_Q A_\mu{}^m \A = \A - i \bar\varepsilon \gamma_\mu \psi^a
e_a{}^m - i \bar\varepsilon \gamma^a \psi_\mu e_a{}^m, \nonu
\delta_Q e_m{}^a \A = \A - i \bar\varepsilon \gamma^a
\psi_b e_m{}^b, \nonu
\delta_Q \psi_\mu \A = \A 2 D_\mu \varepsilon
+ {D-2 \over (d-2)(D-1)} S \gamma_\mu \varepsilon \nonu
\A \A + {1 \over 2} F_{\alpha\beta}^m e_{mc} \left(
e_\mu{}^\alpha \gamma^\beta  - {1 \over 2(d-2)} \gamma_\mu
\gamma^{\alpha\beta} \right) \gamma^c \varepsilon + O(\psi^2), \nonu
\delta_Q \psi_a \A = \A P_{\beta ac} \gamma^c \gamma^\beta \varepsilon
+ {D-2 \over (d-2)(D-1)} S \gamma_a \varepsilon \nonu
\A \A - {1 \over 4} \left( \delta_a^b + {1 \over d-2} \gamma_a \gamma^b
\right) F_{\beta\gamma}^m e_{mb} \gamma^{\beta\gamma} \varepsilon
+ O(\psi^2), \nonu
\delta_Q S \A = \A {1 \over 2} i S \bar\varepsilon \gamma^\alpha
\psi_\alpha - {1 \over 2} i \bar\varepsilon \gamma^{\alpha\beta}
\psi_{\alpha\beta} + {1 \over D-2} i \bar\varepsilon \gamma^\alpha
\gamma^b D_\alpha \psi_b \nonu
\A \A + {1 \over 8} i F_{\alpha\beta}^m e_{mc} \left(
\bar\varepsilon \gamma^{\alpha\beta\delta} \gamma^c \psi_\delta
- {D-3 \over D-2} \bar\varepsilon \gamma^{\alpha\beta} \psi^c
+ {1 \over D-2} \bar\varepsilon \gamma^{\alpha\beta}
\gamma^{cd} \psi_d \right) \nonu
\A \A - {1 \over 2(D-2)} \, i P_{\alpha c}{}^c \bar\varepsilon
\gamma^\beta \gamma^\alpha \psi_\beta
- {1 \over 2} \, i P_{\alpha bc} \bar\varepsilon \gamma^\alpha
\gamma^b \psi^c \nonu
\A \A + {1 \over 2(D-2)} \, i P_{\alpha c}{}^c \bar\varepsilon
\gamma^\alpha \gamma^b \psi_b + O(\psi^3),
\label{b8}
\ea
where the covariant derivative of the parameter is
\be
D_\mu \varepsilon = \left( \partial_\mu
+ {1 \over 4} \omega_\mu{}^{\alpha\beta} \gamma_{\alpha\beta}
+ {1 \over 4} Q_\mu{}^{ab} \gamma_{ab} \right) \varepsilon.
\label{b9}
\ee
%
%%%%%%%  Appendix C  %%%%%%%%%%%%%%%%%%%%%%%%%%%%%%%%%%%%%%%%%%
%
\def\numberbysectionc{\@addtoreset{equation}{section}
\def\theequation{C.\arabic{equation}}}
\numberbysectionc
\vspace{7mm}
\pagebreak[3]
\setcounter{section}{1}
\setcounter{equation}{0}
\setcounter{subsection}{0}
\setcounter{footnote}{0}
\begin{center}
{\large{\bf Appendix C. Dimensional reduction to two dimensions}}
\end{center}
\nopagebreak
\medskip
\nopagebreak
\hspace{3mm}
A dimensional reduction from $D = 3$ to two dimensions requires a
special treatment. The resulting theory in two dimensions turns out
to be a supersymmetric version of dilaton gravity \cite{CHAM}.
We obtain the supertransformations and identify
two-dimensional supergravity multiplet and a matter supermultiplet.
\par
Since eq.\ (\ref{six1}) is singular for $d = 2$,
we use a different parametrization without (super) Weyl rescalings.
Three-dimensional fields $E_M{}^A, \Psi_M, \hat S$
are parametrized in terms of two-dimensional fields
$e_\mu{}^\alpha, A_\mu, X, \psi_\mu, \lambda, S$ as
\ba
E_M{}^A \A = \A \left(
\begin{array}{ccc}
e_\mu{}^\alpha & A_\mu{} \e^{-X} \\
0              & \e^{-X}
\end{array}
\right), \nonu
E_\alpha{}^M \Psi_M \A = \A e_\alpha{}^\mu \psi_\mu, \quad
E_2{}^M \Psi_M = \gamma_2 \lambda, \nonu
\hat S \A = \A 2 S - F' + i \bar\lambda \lambda,
\label{c1}
\ea
where $F'$ is the supercovariantized field strength
\be
F' = - {1 \over 2} \epsilon^{\alpha\beta}
\left( F_{\alpha\beta} \e^{-X}
+ i \bar\psi_\alpha \gamma_\beta \gamma_2 \lambda
+ {1 \over 2} i \bar\psi_\alpha \gamma_2 \psi_\beta \right).
\label{c2}
\ee
\par
Two-dimensional supertransformations are defined by
\ba
\delta_Q (\varepsilon) \A = \A
\delta_{\hat Q}(\varepsilon)
+ \delta_{\hat L}(\lambda(\varepsilon)), \nonu
\lambda_{\alpha 2}(\varepsilon) \A = \A
i \bar\varepsilon \gamma_\alpha \gamma_2 \lambda, \quad
\lambda_{\alpha\beta}(\varepsilon) = 0,
\label{c3}
\ea
where a local Lorentz transformation is added to preserve the
condition $E_m{}^\alpha = 0$.
We find the supertransformations of the fields as
\ba
\delta_Q e_\mu{}^\alpha \A = \A
- i \bar\varepsilon \gamma^\alpha \psi_\mu, \nonu
\delta_Q \psi_\mu \A = \A
2 \left( D_\mu + {1 \over 2} S \gamma_\mu \right) \varepsilon, \nonu
\delta_Q S \A = \A
{1 \over 2} i S \bar\varepsilon \gamma^\mu \psi_\mu
- {1 \over 2} i \bar\varepsilon \gamma^{\mu\nu} \psi_{\mu\nu}, \nonu
\delta_Q X \A = \A i \bar\varepsilon \lambda, \nonu
\delta_Q A_\mu \A = \A
- i \bar\varepsilon \gamma_\mu \gamma_2 \lambda \e^X
- i \bar\varepsilon \gamma_2 \psi_\mu \e^X, \nonu
\delta_Q \lambda \A = \A - \gamma^\mu \varepsilon D^P_\mu X
- \varepsilon \left( F' - S - {1 \over 2} i
\bar\lambda \lambda \right),
\label{c4}
\ea
where $D^P_\mu$ is the supercovariant derivative
(\ref{supercovariantder}).
To obtain the transformation of $S$ we have used
\ba
\delta_Q F' \A = \A
{1 \over 2} i F' \bar\varepsilon \gamma^\mu \psi_\mu
- i \left( F' - S \right)
\bar\varepsilon \lambda
- {1 \over 2} i \bar\varepsilon \gamma^{\mu\nu} \psi_{\mu\nu}
+ i \bar\varepsilon \gamma^\mu D_\mu \lambda \nonu
\A \A + {1 \over 2} i D^P_\mu X \bar\varepsilon
\gamma^\nu \gamma^\mu \psi_\nu
+ i D^P_\mu X \bar\varepsilon \gamma^\mu \lambda
+ {1 \over 4} \bar\varepsilon \gamma^\mu \psi_\mu
\bar\lambda \lambda.
\label{c5}
\ea
Comparing eq.\ (\ref{c4}) with eq.\ (\ref{supertrans}),
we find that the fields $e_\mu{}^\alpha, \psi_\mu, S$
transform as the two-dimensional supergravity multiplet.
The other fields $X, \lambda, A_\mu$ form a matter supermultiplet.
This multiplet is similar to the scalar supermultiplet in
eq.\ (\ref{supertrmatter}) but contains a vector field $A_\mu$
instead of a scalar auxiliary field $F$. Notice that the
fields $A_\mu$ and $S$ have the same off-shell degrees of freedom.
Actually we can relate these two supermultiplets.
If we define
\be
F = F' - S - {1 \over 2} i \bar\lambda \lambda,
\label{c6}
\ee
the supertransformations of $X, \lambda, F$ derived from
eq.\ (\ref{c4}) become exactly the same as
eq.\ (\ref{supertrmatter}) with $d = 2$.
\par
Two-dimensional action dimensionally reduced from the
three-dimensional action (\ref{supergravityaction}) with $d = 3$
is found to be
\be
S_{\rm SG} = {1 \over 16\pi G_0} \int d^2 x \,
e \e^{-X} \left[ R + 2 F S - i S \bar\lambda \lambda
- i \bar\lambda \gamma^{\mu\nu} \psi_{\mu\nu} \right].
\label{c7}
\ee
This is the action of a supersymmetric dilaton gravity \cite{CHAM}.
Strictly speaking, our theory is not exactly the same as the
theories in ref.\ \cite{CHAM}, which use a scalar supermultiplet
with a fundamental scalar auxiliary field.
In the action (\ref{c7}) the field $F$ is not fundamental
but is constructed from other fields as in eq.\ (\ref{c6}).
%
%%%%%%%  References  %%%%%%%%%%%%%%%%%%%%%%%%%%%%%%%%%%%%%%%
\newpage
\newcommand{\NP}[1]{{\it Nucl.\ Phys.\ }{\bf #1}}
\newcommand{\PL}[1]{{\it Phys.\ Lett.\ }{\bf #1}}
\newcommand{\CMP}[1]{{\it Commun.\ Math.\ Phys.\ }{\bf #1}}
\newcommand{\MPL}[1]{{\it Mod.\ Phys.\ Lett.\ }{\bf #1}}
\newcommand{\IJMP}[1]{{\it Int.\ J.\ Mod.\ Phys.\ }{\bf #1}}
\newcommand{\PR}[1]{{\it Phys.\ Rev.\ }{\bf #1}}
\newcommand{\PRL}[1]{{\it Phys.\ Rev.\ Lett.\ }{\bf #1}}
\newcommand{\PTP}[1]{{\it Prog.\ Theor.\ Phys.\ }{\bf #1}}
\newcommand{\PTPS}[1]{{\it Prog.\ Theor.\ Phys.\ Suppl.\ }{\bf #1}}
\newcommand{\AP}[1]{{\it Ann.\ Phys.\ }{\bf #1}}
\newcommand{\JP}[1]{{\it J. Phys.\ }{\bf #1}}
\newcommand{\ZP}[1]{{\it Z. Phys.\ }{\bf #1}}
\newcommand{\NCL}[1]{{\it Nuovo Cimento Lett.\ }{\bf #1}}
\end{document}